\documentclass[fleqn,usenatbib,useAMS]{mnras}
\usepackage{newtxtext,newtxmath}

\usepackage{graphicx}	% Including figure files
\usepackage{amsmath}	% Advanced maths commands
\usepackage{multicol}        % Multi-column entries in tables
\usepackage{bm}		% Bold maths symbols, including upright Greek
\usepackage{pdflscape}	% Landscape pages

\usepackage{tikz}
\usetikzlibrary{calc, positioning}

\usepackage[percent]{overpic}

\usepackage{booktabs, multirow, makecell}

\setcellgapes{5pt}
\newcommand{\vect}[1]{\boldsymbol{#1}}

\newcommand{\Ms}{ \mathrm{M}_\odot }

\newcommand{\Mr}{\frac{\mathrm{M31}}{\mathrm{MW}}}
\newcommand{\LL}{ \mathrm{L} }

\newcommand{\lc}{\lambda_{\mathrm{cut}}} %lambda_cut

\title[The SIBELIUS Project: E Pluribus unum]{The SIBELIUS Project: E Pluribus Unum}

\author[T. Sawala et al.]{Till Sawala,$^{1}$\thanks{E-mail: till.sawala@helsinki.fi}, Stuart McAlpine$^{1}$, Jens Jasche$^{2},$
Guilhem Lavaux$^{3}$, \newauthor Adrian~Jenkins$^{4}$, Peter~H.~Johansson$^{1}$,  Carlos~S.~Frenk$^{4}$ \\
$^{1}$Department of Physics, Gustaf H\"allstr\"omin katu 2, University of Helsinki, Finland\\
$^{2}$The Oskar Klein Centre, Department of Physics, Stockholm University, Albanova University Center, 106 91 Stockholm, Sweden \\
$^{3}$CNRS \& Sorbonne Universit\'e, UMR7095, Institut d'Astrophysique de Paris, 75014 Paris, France\\
$^{4}$Institute for Computational Cosmology, Durham University, South Road, Durham DH1 3LE, United Kingdom \\
}

\date{Accepted XXX. Received YYY; in original form ZZZ}
\pubyear{2019}

\begin{document}
\label{firstpage}
\pagerange{\pageref{firstpage}--\pageref{lastpage}}
\maketitle

% Abstract of the paper
\begin{abstract}
We introduce "Simulations Beyond The Local Universe" (SIBELIUS) that connect the Local Group to its cosmic environment. We show that introducing hierarchical small-scale perturbations to a density field constrained on large scales by observations provides an efficient way to explore the sample space of Local Group analogues. From more than 60~000 simulations, we identify a hierarchy of Local Group characteristics emanating from different scales: the total mass, orientation, orbital energy and the angular momentum are largely determined by modes above $\lambda = 1.6$ comoving Mpc (cMpc) in the primordial density field. Smaller scale variations are mostly manifest as perturbations to the MW-M31 orbit, and we find that the observables commonly used to describe the Local Group --the MW-M31 separation and radial velocity-- are transient and depend on specifying scales down to 0.2~cMpc in the primordial density field. We further find that the presence of M33/LMC analogues significantly affects the MW-M31 orbit and its sensitivity to small-scale perturbations. We construct initial conditions that lead to the formation of a Local Group whose primary observables precisely match the current observations.
\end{abstract}

% Select between one and six entries from the list of approved keywords.
% Don't make up new ones.

%%%%%%%%%%%%%%%%%%%%%%%%%%%%%%%%%%%%%%%%%%%%%%%%%%

%%%%%%%%%%%%%%%%% BODY OF PAPER %%%%%%%%%%%%%%%%%%

\section{Introduction}
The Local Group, comprising our own Milky Way (MW), the neighbouring Andromeda galaxy (M31), the Large Magellanic Cloud (LMC) and M33, and at least a hundred lower mass galaxies \citep[e.g.][]{Karachentsev-2004, McConnachie-2012, Dooley-2017}, has been the place of important discoveries throughout the history of modern cosmology. The correct understanding of galaxies as distinct stellar systems was first proposed in the early eighteenth century, by Immanuel Kant and Thomas Wright of Durham, in reference to M31 \citep[e.g.][]{Whitrow-1967}, and finally settled in the 1920s by measurements of the distances to M33 and M31 \citep{Hubble-1927, Hubble-1929}. In the 1970s, the measurement of the HI rotation curve of M31 suggested the existence of dark matter inside galaxies \citep{Rubin-1970}, and by the 1980s, the velocity dispersions of Local Group dwarf spheroidal galaxies indicated exceptionally high dark matter fractions \citep{Faber-1983}.

Today the Local Group (LG) continues to be one of the most sensitive laboratories to study astrophysical phenomena \citep{Mateo-1998} and cosmological models. Its assembly history, dynamical evolution, and total mass are linked to important open questions in cosmology. In recent years, semi-analytical models \citep[e.g.][]{Bullock-2001, Benson-2002, Somerville-2002} and hydrodynamical simulations that include baryonic physics \citep[e.g.][]{Okamoto-2008, Sawala-2013, Sawala-2014b, Garrison-Kimmel-2019} have shown how the combined effects of reionisation and supernova feedback suppress star formation in, and even the growth of, low mass dark matter haloes. At the same time, the rapid ejection of interstellar gas from the centre of low-mass haloes following bursts of star formation may explain the transformation from a steep, NFW-like cusp \citep{NFW-1996} into a shallower dark matter density profile \citep{Eke-1998, Read-2005, Governato-2010, Teyssier-2013}. While some of the astrophysical processes are still not completely understood, let alone numerically resolved, these baryonic effects have at least offered plausible explanations to some of the challenges raised to the $\Lambda$CDM model \citep{Moore-1999, Boylan-Kolchin-2011}. However, the uncertainty in the total mass of the Local Group of at least a factor of two \citep{Li-2008, vanderMarel-2012} remains a major caveat for all predictions of the amount of substructure.

In the meantime, another question that has been asked since at least the seventeenth century, but which dates back to the Copernican revolution of the sixteenth century, has become more pertinent: how special is our own vantage point on the Universe, and what can we induce about cosmology by closely studying the finite patch of the Universe that surrounds us? Even more pressing, perhaps, is the inverse: how plausible is the formation of the Local Group in the assumed cosmological model? It has recently been pointed out that the LG has characteristics which make it exceedingly rare in $\Lambda$CDM: only a small fraction of Milky Way sized haloes in $\Lambda$CDM simulations have satellites as massive as the LMC and SMC~\citep{Benson-2002, Boylan-Kolchin-2011, Tollerud-2011, Robotham-2012}, the measured MW-M31 orbit is unusually radial \cite[e.g.][]{Fattahi-2015}, and only a very small fraction of $\Lambda$CDM haloes have satellite systems as anisotropic as the Local Group's \citep[e.g.][]{Ibata-2013, Pawlowski-2015}.

The above mentioned baryonic physics seem to offer no possible path to answer these questions~\citep{Bullock-2017, Pawlowski-review}. Furthermore, "simple" modifications or extensions to the $\Lambda$CDM model, such as warm dark matter \citep[e.g.][]{Lovell-2011}, a modification of the primordial power spectrum~\citep[e.g.][]{Enqvist-2020}, or self-interacting DM~\citep[e.g.][]{Lovell-2020} are no convincing alternatives. If the particular features of the Local Universe are as exceptional as they are claimed to be, they may hint at changes to our cosmological models that go beyond what has been so far envisioned. Alternatively, they should lead to an understanding of the unique but perhaps not implausible environment in which the Local Universe has evolved.

The aim of the {\sc Sibelius} (Simulations Beyond the Local Universe) project is to elucidate the co-evolution of the Local Group, the Local Universe, and the large scale structure (LSS). Similar constrained local universe simulations include the {\sc Clues} \citep{Gottloeber-2010, Libeskind-2010, Carlesi-2016} and {\sc Hestia} \citep{Libeskind-2020} projects, which embed several Local Group analogues within the observed large scale structure. However, our method allows us to systematically examine a much larger set of initial conditions, to not only create a precise match to the Local Group's orbit, but to also to reveal the relation between its parameters, and their dependence on the initial density field.

In this paper, we present our new method for creating initial conditions that reproduce the kinematics of the Local Group, and describe the sensitivity of the LG observables to different scales in the initial conditions.  We make extensive use of the methods described in ~\citet{Sawala-2020} to create a series of hierarchical, random perturbation of the primordial density field to unveil the scales at which the observed properties of the Local Group are set, and identify the properties that characterise it.

This paper is organised as follows. In Section~\ref{sec:methods}, we describe the methods: the parametrisation of the phase information (\ref{sec:methods:phase}), the large scale structure constraints (\ref{sec:methods:constraints}), their octree function representation (\ref{sec:methods:octree_lss}), and the implementation of multi-level random variations (\ref{sec:methods:variations}). We describe the numerical simulations used (\ref{sec:methods:simulations}) and the identification of LG analogues (\ref{sec:methods:identification}). Sections~\ref{sec:results:firstorder} to~\ref{sec:higher-order} present the results. In Section~\ref{sec:numbers}, we discuss the number of LG analogues identified when all unconstrained scales are varied, and in Section~\ref{sec:multiplicity}, we discuss multiple occurrences of the same object in our large random sample. In Section~\ref{sec:quality-first-order}, we introduce a metric to compare LG analogues to observations. In Section~\ref{sec:characteristics}, we identify the "characteristics" of the Local Group as those quantities that are defined on the largest scales. In Section~\ref{sec:third-obects}, we discuss the frequency of LMC/M33 analogues. In Section~\ref{sec:higher-order}, we discuss smaller scale variation, and determine in~\ref{sec:orbital-phase} the scale that sets the LG's orbital phase.

\begin{figure}

\resizebox{!}{6.5cm}{%
\begin{tikzpicture}[scale=.9,every node/.style={minimum size=1cm},on grid]

    \draw
    (-3.5,3.) node {\Large L=18}
    (-3.5,4.4) node {\Large L=19}
    (-3.5,5.8) node {\Large L=20}
    (-3.5,7.2) node {\Large L=21}
    (-3.5,8.6) node {\Large L=22}
    (-3.5,10.) node {\Large L=23}
    (-3.5,11.4) node {\Large L=24}
    (-3.5,12.8) node {\Large L=25};

     \draw
    (0.,15.) node {\LARGE Constrained Simulation};
    
    \begin{scope}[yshift=50,every node/.append style={yslant=0.5,xslant=-1,},yslant=0.5,xslant=-1]
        \fill[white,fill opacity=.9] (0,0) rectangle (2.5,2.5);
        \draw[step=1.25, black] (0,0) grid (2.5,2.5);
        \draw[black,very thick] (0,0) rectangle (2.5,2.5);
        \fill[blue, fill opacity = 0.1] (0,0) rectangle (2.5,2.5);
    \end{scope}
    
    % constraint, 19
    \begin{scope}[yshift=90,every node/.append style={yslant=0.5,xslant=-1},yslant=0.5,xslant=-1]
    	\fill[white,fill opacity=.9] (0,0) rectangle (2.5,2.5);
        \draw[black,very thick] (0,0) rectangle (2.5,2.5);
        \draw[step=0.625, black] (0,0) grid (2.5,2.5);
        \fill[blue, fill opacity = 0.1] (0,0) rectangle (.625,2.5);
        \fill[blue, fill opacity = 0.1] (.625,1.875) rectangle (1.875,2.5);
        \fill[blue, fill opacity = 0.1] (.625,0) rectangle (1.875,.625);
        \fill[blue, fill opacity = 0.1] (1.875,0) rectangle (2.5,2.5);
        \fill[blue, fill opacity = 0.3] (.625,.625) rectangle (1.875,1.875);
    \end{scope}  
    
    % constraint, 20
    \begin{scope}[yshift=130,every node/.append style={yslant=0.5,xslant=-1},yslant=0.5,xslant=-1]
    	\fill[white,fill opacity=.9] (0,0) rectangle (2.5,2.5);
        \draw[black,very thick] (0,0) rectangle (2.5,2.5);
        \draw[step=0.3125, black] (0,0) grid (2.5,2.5);
        \fill[blue, fill opacity = 0.1] (0,0) rectangle (.625,2.5);
        \fill[blue, fill opacity = 0.1] (.625,1.875) rectangle (1.875,2.5);
        \fill[blue, fill opacity = 0.1] (.625,0) rectangle (1.875,.625);
        \fill[blue, fill opacity = 0.1] (1.875,0) rectangle (2.5,2.5);
        \fill[blue, fill opacity = 0.3] (.625,.625) rectangle (1.875,1.875);
    \end{scope}

    % constraint, 21
   \begin{scope}[yshift=170,every node/.append style={yslant=0.5,xslant=-1},yslant=0.5,xslant=-1]
    	\fill[white,fill opacity=.9] (0,0) rectangle (2.5,2.5);
        \draw[black,very thick] (0,0) rectangle (2.5,2.5);
        \draw[step=0.15625, black] (0,0) grid (2.5,2.5);
          \fill[blue, fill opacity = 0.1] (0,0) rectangle (.625,2.5);
        \fill[blue, fill opacity = 0.1] (.625,1.875) rectangle (1.875,2.5);
        \fill[blue, fill opacity = 0.1] (.625,0) rectangle (1.875,.625);
        \fill[blue, fill opacity = 0.1] (1.875,0) rectangle (2.5,2.5);
        \fill[blue, fill opacity = 0.3] (.625,.625) rectangle (1.875,1.875);
    \end{scope}
    
        % unconstrained, 22
    \begin{scope}[yshift=210,every node/.append style={yslant=0.5,xslant=-1},yslant=0.5,xslant=-1]
    	\fill[white,fill opacity=.9] (0,0) rectangle (2.5,2.5);
        \draw[black,very thick] (0,0) rectangle (2.5,2.5);
        \draw[step=0.078125, black] (0,0) grid (2.5,2.5);
        \fill[yellow, fill opacity = 0.3] (0,0) rectangle (2.5,2.5);
    \end{scope}
    
\end{tikzpicture}
} % end the resizebox
\resizebox{!}{6.5cm}{
\begin{tikzpicture}[scale=.9,every node/.style={minimum size=1cm},on grid]
       \draw
    (0.,15.) node {\LARGE 3.2~cMpc Variants};
    
    \begin{scope}[yshift=50,every node/.append style={yslant=0.5,xslant=-1,},yslant=0.5,xslant=-1]
        \fill[white,fill opacity=.9] (0,0) rectangle (2.5,2.5);
        \draw[step=1.25, black] (0,0) grid (2.5,2.5);
        \draw[black,very thick] (0,0) rectangle (2.5,2.5);
        \fill[blue, fill opacity = 0.1] (0,0) rectangle (2.5,2.5);
    \end{scope}
    
    % constraint, 19
    \begin{scope}[yshift=90,every node/.append style={yslant=0.5,xslant=-1},yslant=0.5,xslant=-1]
    	\fill[white,fill opacity=.9] (0,0) rectangle (2.5,2.5);
        \draw[black,very thick] (0,0) rectangle (2.5,2.5);
        \draw[step=0.625, black] (0,0) grid (2.5,2.5);
        \fill[blue, fill opacity = 0.1] (0,0) rectangle (.625,2.5);
        \fill[blue, fill opacity = 0.1] (.625,1.875) rectangle (1.875,2.5);
        \fill[blue, fill opacity = 0.1] (.625,0) rectangle (1.875,.625);
        \fill[blue, fill opacity = 0.1] (1.875,0) rectangle (2.5,2.5);
        \fill[blue, fill opacity = 0.3] (.625,.625) rectangle (1.875,1.875);
    \end{scope}  
    
    % constraint, 20
    \begin{scope}[yshift=130,every node/.append style={yslant=0.5,xslant=-1},yslant=0.5,xslant=-1]
    	\fill[white,fill opacity=.9] (0,0) rectangle (2.5,2.5);
        \draw[black,very thick] (0,0) rectangle (2.5,2.5);
        \draw[step=0.3125, black] (0,0) grid (2.5,2.5);
        \fill[blue, fill opacity = 0.1] (0,0) rectangle (.625,2.5);
        \fill[blue, fill opacity = 0.1] (.625,1.875) rectangle (1.875,2.5);
        \fill[blue, fill opacity = 0.1] (.625,0) rectangle (1.875,.625);
        \fill[blue, fill opacity = 0.1] (1.875,0) rectangle (2.5,2.5);
        \fill[blue, fill opacity = 0.3] (.625,.625) rectangle (1.875,1.875);
    \end{scope}

    % constraint, 21
   \begin{scope}[yshift=170,every node/.append style={yslant=0.5,xslant=-1},yslant=0.5,xslant=-1]
    	\fill[white,fill opacity=.9] (0,0) rectangle (2.5,2.5);
        \draw[black,very thick] (0,0) rectangle (2.5,2.5);
        \draw[step=0.15625, black] (0,0) grid (2.5,2.5);
          \fill[blue, fill opacity = 0.1] (0,0) rectangle (.625,2.5);
        \fill[blue, fill opacity = 0.1] (.625,1.875) rectangle (1.875,2.5);
        \fill[blue, fill opacity = 0.1] (.625,0) rectangle (1.875,.625);
        \fill[blue, fill opacity = 0.1] (1.875,0) rectangle (2.5,2.5);
        \fill[blue, fill opacity = 0.3] (.625,.625) rectangle (1.875,1.875);
    \end{scope}
    
        % unconstrained, 22
    \begin{scope}[yshift=210,every node/.append style={yslant=0.5,xslant=-1},yslant=0.5,xslant=-1]
    	\fill[white,fill opacity=.9] (0,0) rectangle (2.5,2.5);
        \draw[black,very thick] (0,0) rectangle (2.5,2.5);
        \draw[step=0.078125, black] (0,0) grid (2.5,2.5);
        \fill[red, fill opacity = 0.6] (0,0) rectangle (2.5,2.5);
    \end{scope}
    
    % unconstrained, 23 
    \begin{scope}[yshift=250,every node/.append style={yslant=0.5,xslant=-1},yslant=0.5,xslant=-1]
    	\fill[white,fill opacity=.9] (0,0) rectangle (2.5,2.5);
        \draw[black,very thick] (0,0) rectangle (2.5,2.5);
        \draw[step=0.0390625, black] (0,0) grid (2.5,2.5);
        \fill[red, fill opacity = 0.6] (0,0) rectangle (2.5,2.5);
    \end{scope} 
  
    % unconstrained, 24
    \begin{scope}[yshift=290,every node/.append style={yslant=0.5,xslant=-1},yslant=0.5,xslant=-1]
    	\fill[white,fill opacity=.9] (0,0) rectangle (2.5,2.5);
        \draw[black,very thick] (0,0) rectangle (2.5,2.5);
        \fill[red, fill opacity = 0.6] (0,0) rectangle (2.5,2.5);
    \end{scope} 
  \end{tikzpicture}
} % end the resizebox
\resizebox{!}{6.5cm}{
\begin{tikzpicture}[scale=.9,every node/.style={minimum size=1cm},on grid]

       \draw
    (0.,15.) node {\LARGE 0.8~cMpc Variants};
    
    %level 18
    \begin{scope}[yshift=50,every node/.append style={yslant=0.5,xslant=-1,},yslant=0.5,xslant=-1]
        \fill[white,fill opacity=.9] (0,0) rectangle (2.5,2.5);
        \draw[step=1.25, black] (0,0) grid (2.5,2.5);
        \draw[black,very thick] (0,0) rectangle (2.5,2.5);
        \fill[blue, fill opacity = 0.1] (0,0) rectangle (2.5,2.5);
    \end{scope}
    
    % constraint, 19
    \begin{scope}[yshift=90,every node/.append style={yslant=0.5,xslant=-1},yslant=0.5,xslant=-1]
    	\fill[white,fill opacity=.9] (0,0) rectangle (2.5,2.5);
        \draw[black,very thick] (0,0) rectangle (2.5,2.5);
        \draw[step=0.625, black] (0,0) grid (2.5,2.5);
        \fill[blue, fill opacity = 0.1] (0,0) rectangle (.625,2.5);
        \fill[blue, fill opacity = 0.1] (.625,1.875) rectangle (1.875,2.5);
        \fill[blue, fill opacity = 0.1] (.625,0) rectangle (1.875,.625);
        \fill[blue, fill opacity = 0.1] (1.875,0) rectangle (2.5,2.5);
        \fill[blue, fill opacity = 0.3] (.625,.625) rectangle (1.875,1.875);
    \end{scope}  
    
    % constraint, 20
    \begin{scope}[yshift=130,every node/.append style={yslant=0.5,xslant=-1},yslant=0.5,xslant=-1]
    	\fill[white,fill opacity=.9] (0,0) rectangle (2.5,2.5);
        \draw[black,very thick] (0,0) rectangle (2.5,2.5);
        \draw[step=0.3125, black] (0,0) grid (2.5,2.5);
        \fill[blue, fill opacity = 0.1] (0,0) rectangle (.625,2.5);
        \fill[blue, fill opacity = 0.1] (.625,1.875) rectangle (1.875,2.5);
        \fill[blue, fill opacity = 0.1] (.625,0) rectangle (1.875,.625);
        \fill[blue, fill opacity = 0.1] (1.875,0) rectangle (2.5,2.5);
        \fill[blue, fill opacity = 0.3] (.625,.625) rectangle (1.875,1.875);
    \end{scope}

    % constraint, 21
   \begin{scope}[yshift=170,every node/.append style={yslant=0.5,xslant=-1},yslant=0.5,xslant=-1]
    	\fill[white,fill opacity=.9] (0,0) rectangle (2.5,2.5);
        \draw[black,very thick] (0,0) rectangle (2.5,2.5);
        \draw[step=0.15625, black] (0,0) grid (2.5,2.5);
        \fill[blue, fill opacity = 0.1] (0,0) rectangle (.625,2.5);
        \fill[blue, fill opacity = 0.1] (.625,1.875) rectangle (1.875,2.5);
        \fill[blue, fill opacity = 0.1] (.625,0) rectangle (1.875,.625);
        \fill[blue, fill opacity = 0.1] (1.875,0) rectangle (2.5,2.5);
        \fill[blue, fill opacity = 0.3] (.625,.625) rectangle (1.875,1.875);
    \end{scope}
    
        % unconstrained, 22
    \begin{scope}[yshift=210,every node/.append style={yslant=0.5,xslant=-1},yslant=0.5,xslant=-1]
    	\fill[white,fill opacity=.9] (0,0) rectangle (2.5,2.5);
        \draw[black,very thick] (0,0) rectangle (2.5,2.5);
        \draw[step=0.078125, black] (0,0) grid (2.5,2.5);
        \fill[red, fill opacity = 0.1] (0,0) rectangle (2.5,2.5);
    \end{scope}
    
    % unconstrained, 23 
    \begin{scope}[yshift=250,every node/.append style={yslant=0.5,xslant=-1},yslant=0.5,xslant=-1]
    	\fill[white,fill opacity=.9] (0,0) rectangle (2.5,2.5);
        \draw[black,very thick] (0,0) rectangle (2.5,2.5);
        \draw[step=0.0390625, black] (0,0) grid (2.5,2.5);
        \fill[red, fill opacity = 0.1] (0,0) rectangle (2.5,2.5);
    \end{scope} 
  
    % 0.8~cMpc, 24 
    \begin{scope}[yshift=290,every node/.append style={yslant=0.5,xslant=-1},yslant=0.5,xslant=-1]
    	\fill[white,fill opacity=.9] (0,0) rectangle (2.5,2.5);
        \draw[black,very thick] (0,0) rectangle (2.5,2.5);
        \fill[red, fill opacity = 0.1] (0,0) rectangle (1.0,2.5);
        \fill[red, fill opacity = 0.1] (1.5,0) rectangle (2.5,2.5);
        \fill[red, fill opacity = 0.1] (1.0,0) rectangle (1.5,1.0);
        \fill[red, fill opacity = 0.1] (1.0,1.5) rectangle (1.5,2.5);
        \fill[red, fill opacity = 0.6] (1.,1.) rectangle (1.5,1.5);
    \end{scope} 

    % 0.8~cMpc, 25 
    \begin{scope}[yshift=330,every node/.append style={yslant=0.5,xslant=-1},yslant=0.5,xslant=-1]
    	\fill[white,fill opacity=.9] (0,0) rectangle (2.5,2.5);
        \draw[black,very thick] (0,0) rectangle (2.5,2.5);
        \fill[red, fill opacity = 0.1] (0,0) rectangle (1.0,2.5);
        \fill[red, fill opacity = 0.1] (1.5,0) rectangle (2.5,2.5);
        \fill[red, fill opacity = 0.1] (1.0,0) rectangle (1.5,1.0);
        \fill[red, fill opacity = 0.1] (1.0,1.5) rectangle (1.5,2.5);
        \fill[red, fill opacity = 0.6] (1.,1.) rectangle (1.5,1.5);
    \end{scope} 
\end{tikzpicture}
} % end the resizebox
\caption{Illustration of the phase information used in the simulations. For each constrained simulation (shown in Figure~\ref{fig:borg}), the set of constraints sets the coefficients up to level 21, with the remaining levels taken from {\sc Panphasia} (yellow). For the 3.2~cMpc variants (top two rows in Figure~\ref{fig:spheres}), we preserve the constraints, and vary the phase information in the unconstrained levels with different regions from {\sc Panphasia} (red). For each set of 0.8~cMpc variations (bottom two rows in Figure~\ref{fig:spheres}), we also preserve the information from a sample of the 3.2~cMpc variations, and independently randomise a small subregion at levels 24 and 25 (red). Note that, in the actual simulations, the $1000^3$cMpc$^3$ volume is represented by $15^3$ cells at level 18, and the subregions are close to spherical rather than n-cubes. \label{fig:shifts}}
\end{figure}
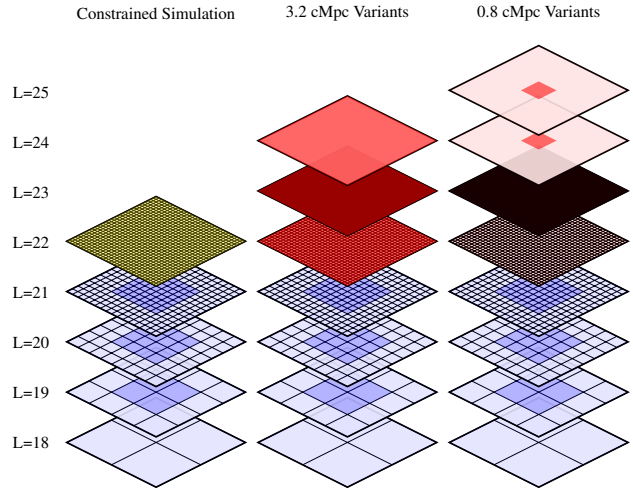

\begin{figure*}
\begin{overpic}[width=1.72in]{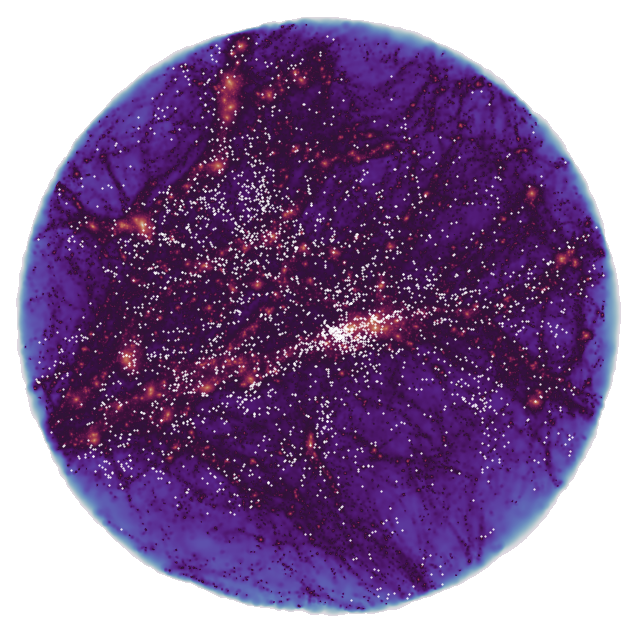} 
 \put (5,90) {\textcolor{black}{\textbf{7600}}}
\end{overpic}
\begin{overpic}[width=1.72in]{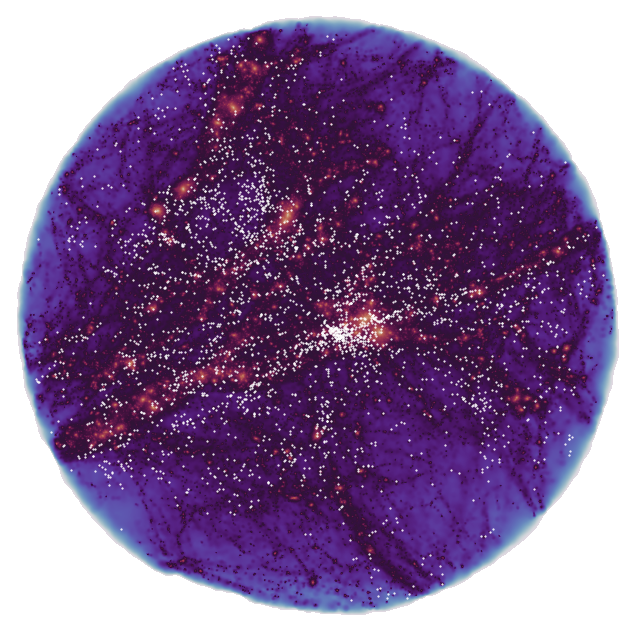} 
 \put (5,90) {\textcolor{black}{\textbf{7850}}}
\end{overpic}
\begin{overpic}[width=1.72in]{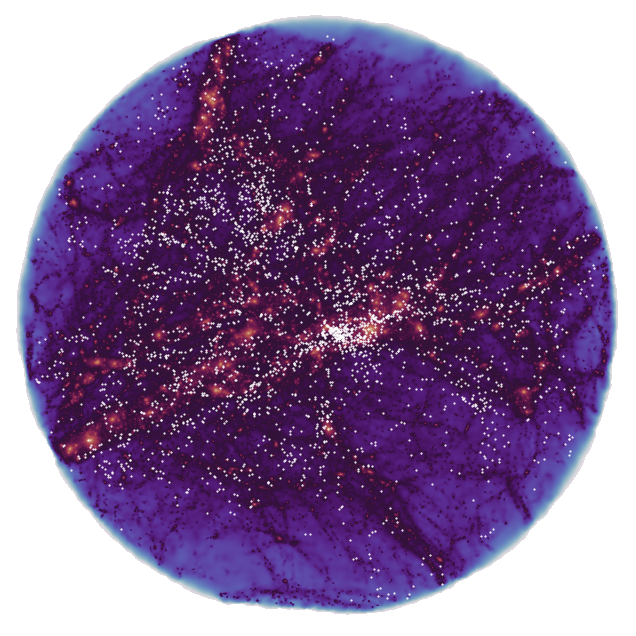} 
 \put (5,90) {\textcolor{black}{\textbf{8100}}}
\end{overpic}
\begin{overpic}[width=1.72in]{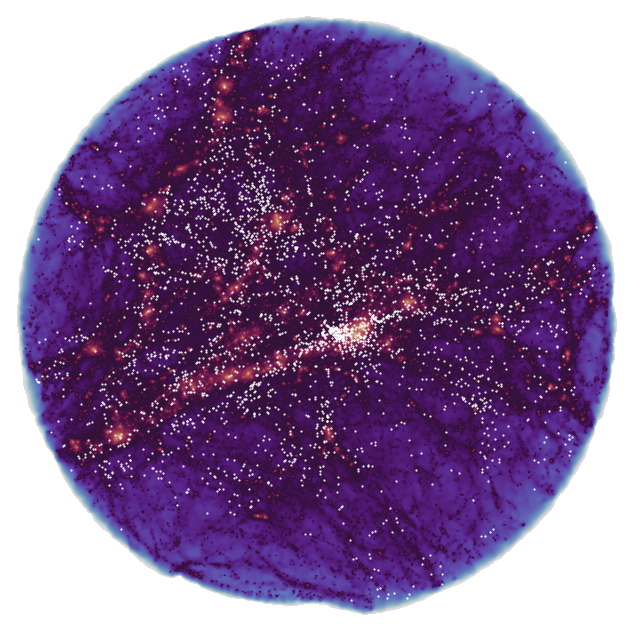} 
 \put (5,90) {\textcolor{black}{\textbf{8350}}}
\end{overpic} \\
\begin{overpic}[width=1.72in]{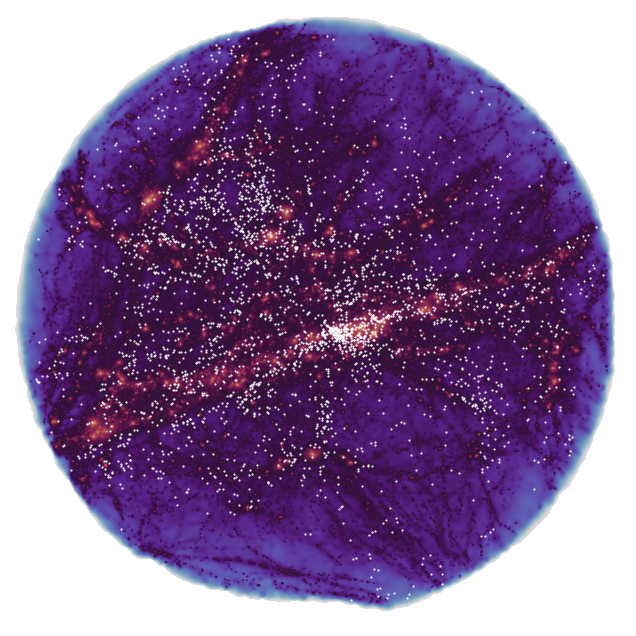} 
 \put (5,90) {\textcolor{black}{\textbf{8600}}}
\end{overpic}
\begin{overpic}[width=1.72in]{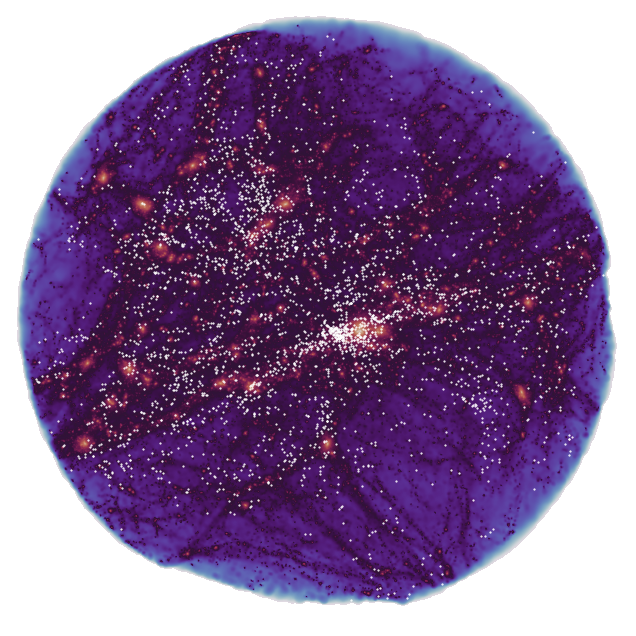} 
 \put (5,90) {\textcolor{black}{\textbf{8850}}}
\end{overpic}
\begin{overpic}[width=1.72in]{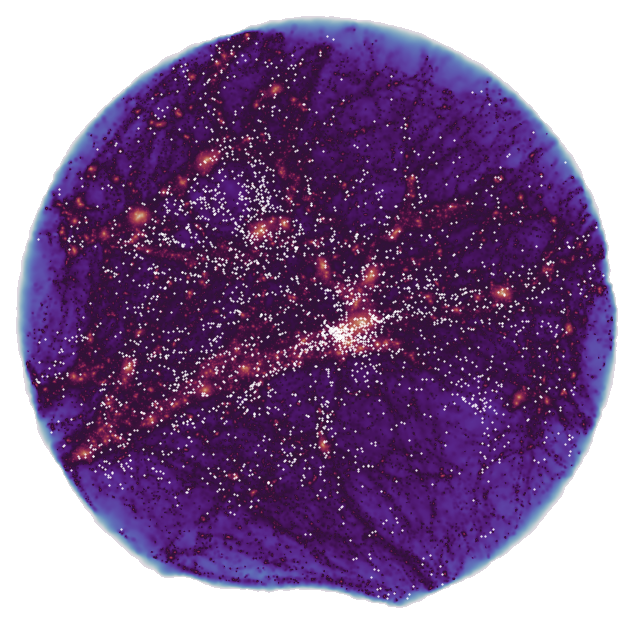} 
 \put (5,90) {\textcolor{black}{\textbf{9100}}}
\end{overpic}
\begin{overpic}[width=1.72in]{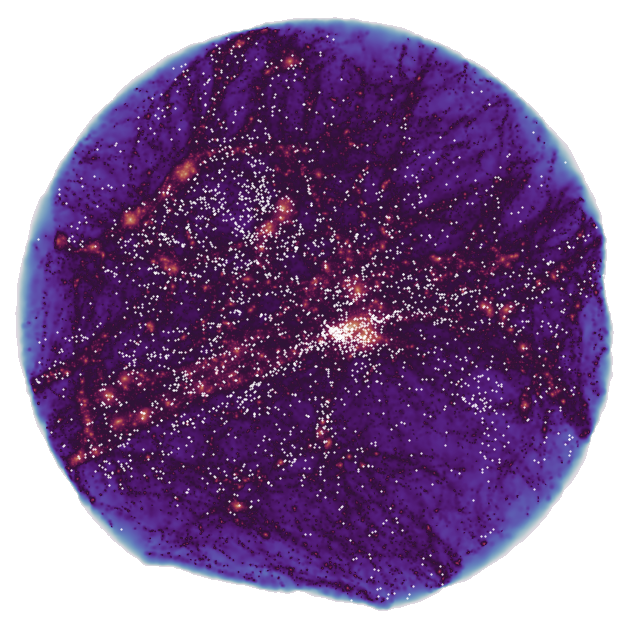} 
 \put (5,90) {\textcolor{black}{\textbf{9350}}}
\end{overpic} 
    \caption{Projected dark matter density in spheres of $r=50$Mpc, centred on the fiducial observer at $z=0$, in labelled iterations of the Markov chain of the {\sc Borg} reconstruction. Overlaid in white are the positions of galaxies in the 2MASS redshift survey \citep{Huchra-12} between $r=10-50$~Mpc. The observed clusters and voids are consistently reproduced, but there are noticeable differences between individual iterations. \label{fig:borg}}
\end{figure*}

\begin{figure}  
\begin{overpic}[width=1.60in]{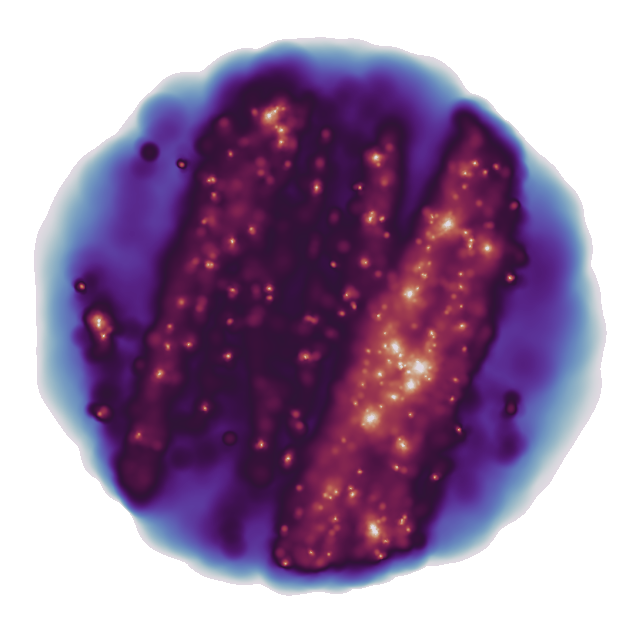}
\linethickness{3pt}    
    \put(77,7) {\textcolor{black}{\bf 2 Mpc}}
    \put(77,3) {\color{black}\line(1,0){20}}
\end{overpic}
    \includegraphics[width=1.60in]{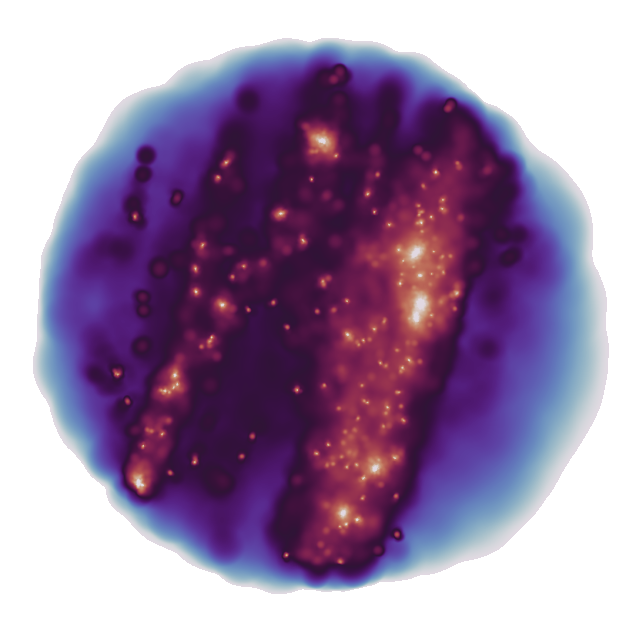} 
    \includegraphics[width=1.60in]{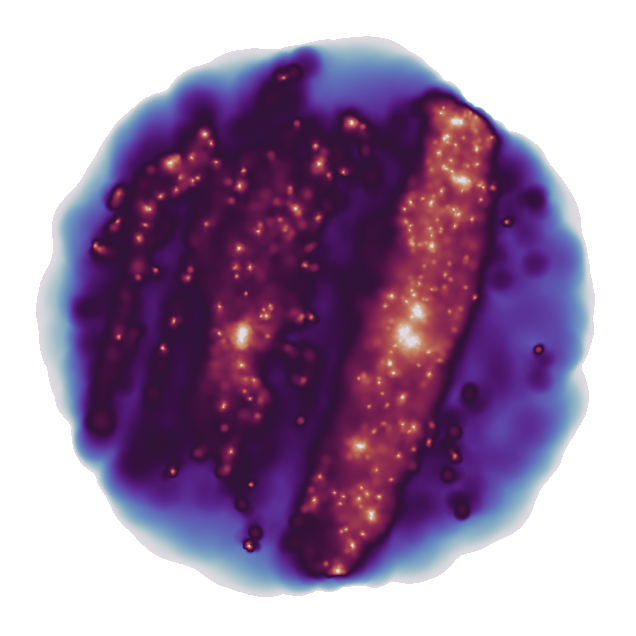}
    \includegraphics[width=1.60in]{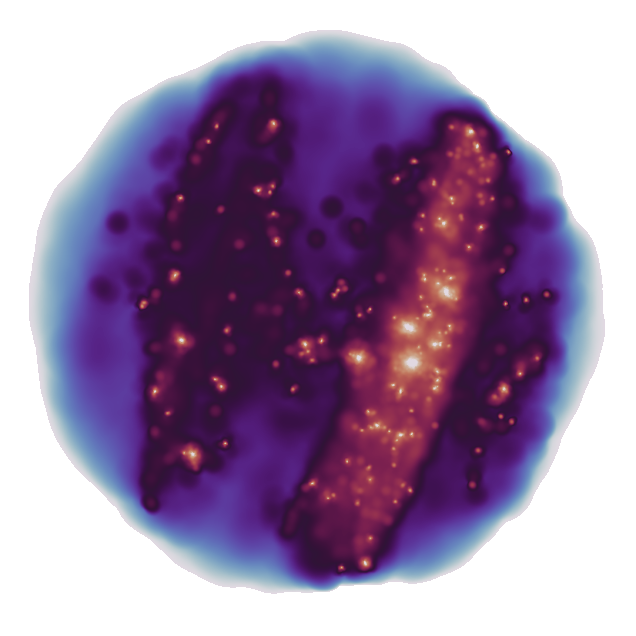}
    
\begin{overpic}[width=1.60in]{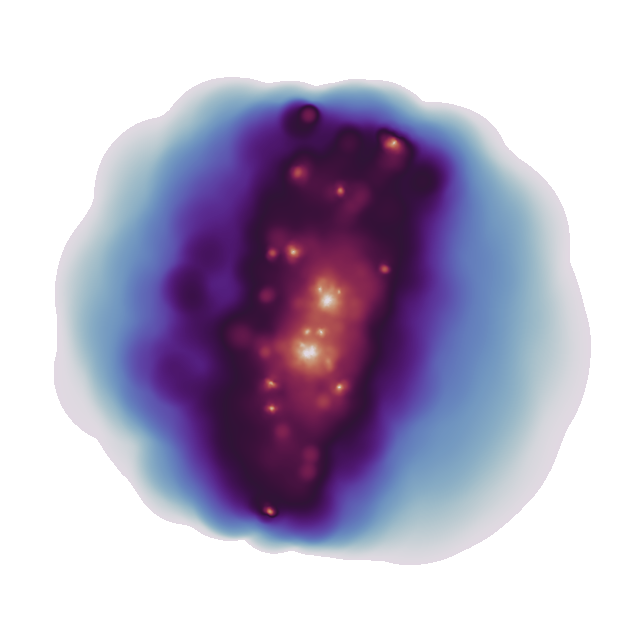}
\linethickness{3pt}    
    \put(77,9) {\textcolor{black}{\bf 1 Mpc}}
    \put(72,5) {\color{black}\line(1,0){25}}
\end{overpic}
    \includegraphics[width=1.60in]{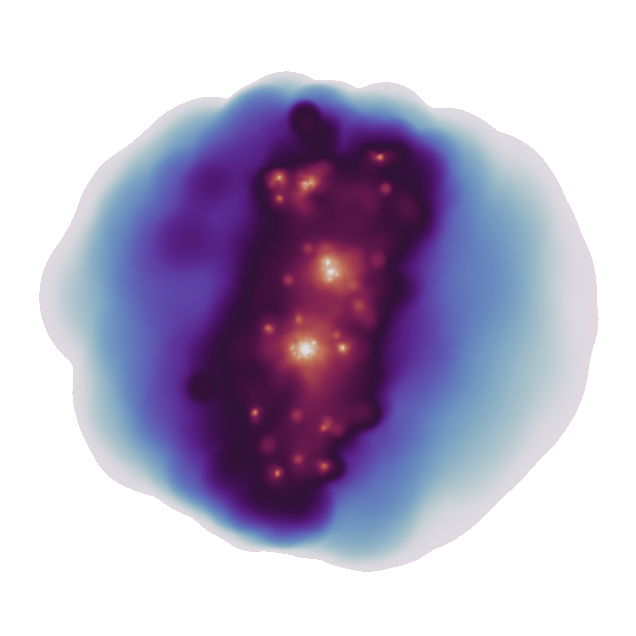} 
    \includegraphics[width=1.60in]{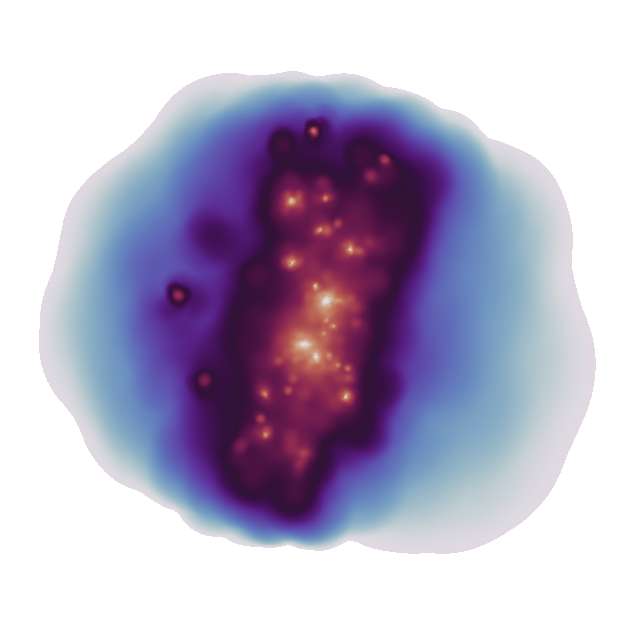}
    \includegraphics[width=1.60in]{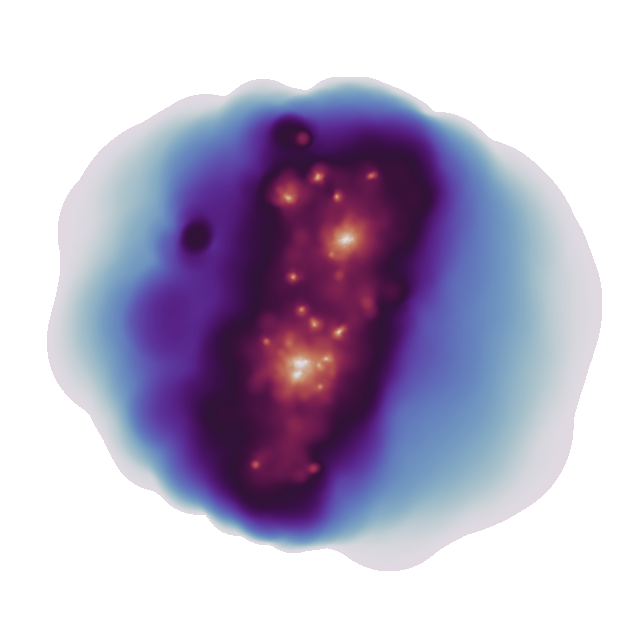}  
    \caption{Top two rows: projected dark matter density in spheres of $r=5$Mpc, centred on the fiducial observer, in four of the 60~000 3.2~cMpc random variations of constraint 9350. Due to the shared large scale modes, all volumes show a similar filamentary structure. However, the identity of individual haloes, determined on smaller scales, changes between volumes. Bottom two rows: spheres of $r=2$Mpc, centred on the fiducial observer, in four of the 0.8~cMpc variations, using the same constraints and the same 3.2~cMpc variation, $i=24493$. Now, all volumes contain a LG analogue of similar mass and orientation, but varying separation. \label{fig:spheres}
    }
\end{figure}

\section{Methods}\label{sec:methods}
All results in this paper are based on Dark Matter Only (DMO) zoom-in simulations using initial conditions generated with the {\sc ic\_gen} code \citep{Jenkins-2010} for a $\Lambda$CDM Universe with the following cosmological parameters: $\Omega_0 = 0.307$, $\Omega_\Lambda = 0.693$, $\sigma_8 = 0.8288$, $h=0.6777$ \citep{Planck-2013}. The initial displacement field is calculated using 2nd-order Lagrangian perturbation theory at $z=169$, although the simulations are not sensitive to the starting redshift. The unperturbed particle positions in the high-resolution region are realised as glasses, with identifiers that encode the initial (unperturbed) position of the particles. We use quadratic interpolation to compute the displacements.

\subsection{Phase Information}\label{sec:methods:phase}
In the $\Lambda$CDM model the primordial density fluctuations after inflation are homogeneous and Gaussian-distributed. To specify the initial conditions for a cosmological $\Lambda$CDM simulation, in addition to the cosmological parameters, we define the dimensions of the periodic volume, the power spectrum, and choose the phases for the realisation. The information describing the particular choices for the phases can be represented by a Gaussian white noise field. Convolving this white noise field in real space with a suitable, real non-negative filter, gives the $\Lambda$CDM 
density field \citep{Salmon-1996}.

In this work we wish to construct a cosmological volume that contains a region resembling the observed large scale structure around us. We use the S$_8$ octree orthogonal basis function set introduced in \cite{Jenkins-2013} as the building blocks for representing the phase information. We choose this representation over the more obvious choice of Fourier modes because the octree representation is designed to be more efficient for building the kind of multi-scale initial conditions we require in the {\sc Sibelius} project. The octree basis functions have compact support and their individual contributions to the $\Lambda$CDM density field are also well localised. This contrasts with Fourier modes which have equal weight everywhere. From a purely computational point of view, using Fourier modes is more expensive when we wish to specify initial conditions at high resolution only within a small region of a large periodic volume.

In Section~\ref{sec:methods:constraints}, we will describe how we use large scale structure constraints to set the amplitudes for some of the physically largest octree functions. On the smallest scales, however, the large scale structure constraints are unimportant and the field is effectively unconstrained. By default, we set the unconstrained coefficients using the Panphasia Gaussian white noise field \citep{Jenkins-2013}. The choice of location within the Panphasia field is arbitrary but by making this choice we determine the physical sizes and locations of the octree cells relative to the simulation volume. We choose to embed the constraints that represent the local large scale structure out to a radius of $\sim 300$~cMpc within a cubic volume of side length 1000 comoving Mpc (hereafter cMpc). We identify this volume with a set of $15^3$ octree cells at  level 18 of the Panphasia octree.

This provides an anchor point for the physical cell size: in all our simulations, the cubic cells at level 18 have a volume of $\sim 66.67^3$~cMpc$^3$; cells at Level $\LL$ have side length $66.67 \times 2^{(18-L)}$~cMpc. In order to connect levels of the octree with a physical scale, in \cite{Sawala-2020}, we also computed $\lc(L)$, the wavelength at which the amplitude of fluctuations of the $\Lambda$CDM power spectrum is reduced by a factor of four when power at levels $\LL$ and above are set to zero. \cite{Jenkins-2013} provides a more in-depth description of the public code {\sc Panphasia}, while \cite{Jenkins-Booth-2013} contains a user guide.

In \cite{Sawala-2020}, we already explored the propagation of changes to the primordial density field to structures at $z=0$. In the present work, we apply this knowledge, first to explore the ensemble of Local Group analogues that can form within the constraints, and then to create variations of individual Local Group analogues by randomising different levels of the white noise field.

As illustrated in Figure~\ref{fig:shifts}, the cosmological constraints described in Section~\ref{sec:methods:constraints} are located on levels 19-21 of the octree, constraining modes down to $\lc(\LL=21) = 6.48$ cMpc. As shown in \cite{Sawala-2020}, this is sufficient to determine the existence of individual $z=0$ haloes above $\sim 10^{14} \Ms$. For haloes of $\sim 10^{15}\Ms$, i.e. the size of the Virgo Cluster, there is an expected residual scatter in position of $\sim 1$~Mpc, and a scatter in mass of $\sim 5\%$, which are both below the observational uncertainties. However, the existence of unique $10^{12}\Ms$ haloes, i.e. in the mass range of both the MW and M31, requires defining the phase information to smaller scales, i.e. to higher levels of the octree. This offers an opportunity: modifications to the phase information on the right scale allows us to create different ensembles of MW / M31 mass haloes without affecting the observed large scale structure.

\subsection{Constraints for the Local Large Scale Structure}\label{sec:methods:constraints}

The phase information for the Sibelius project is created in multiple steps. Our starting point is the phase information obtained using the {\sc Borg} algorithm \citep[Bayesian Origin Reconstruction from Galaxies][]{Jasche-2013,Jasche-2019}, which is represented on a regular $256^3$ mesh covering a volume of $1000^3$~cMpc$^3$, i.e. each volume element covers $\sim 3.9$~cMpc$^3$.

The {\sc Borg} algorithm is designed to infer the primordial density field from input data on the spatial galaxy distribution. Though in principle {\sc Borg} could infer jointly the cosmological parameters, the specific initial conditions used here were derived for a fixed cosmology, which sets a prior on the power spectrum of the primordial density fluctuations. The observational data used to perform the inference is the 2M++ compilation \citep{Lavaux-2011}. The 2M++ galaxies are photometrically selected from the 2MASS sample \citep[2 Micron All Sky Survey][]{Skrutskie-06}. The redshift information has been assembled from the 2MASS Redshift Survey for the brightest galaxies \citep{Huchra-12}, the SDSS DR-7 main galaxy sample \citep[Sloan Digital Sky Survey Data Release 7][]{Abazajian-09}, and the 6dFGRS \citep[6 degree Field Galaxy Redshift Survey][]{Jones-09}. {\sc Borg} is able to account for selection effects; thus there is no sharp boundary to the constrained volume, which in practice reaches out to a distance of about 300~Mpc from the observer. {\sc Borg} also infers self-consistently the bias relation between the galaxy population and the underlying dark matter density field through the use of empirical relations. The details of the inference mechanism are given in \citet{Jasche-2019}.

We note that, by design, {\sc Borg} does not yield a unique result, but instead employs a Markov-chain to represent plausible samples of the probability that each initial condition yields the present universe. Each iteration of the Markov-chain corresponds to a different density field and an associated likelihood value. Even when the overall likelihood converges, the local density field is still subject to significant variations arising from the uncertainties in the properties of the galaxy population in each volume element ($\sim 3.9$ cMpc$^3$). Among the samples with comparable likelihoods, we selected the last iteration (9350) of the chain, which also gave a good reproduction of the masses and positions of the nearby Virgo and Fornax Clusters. A different realisation would lead to differences in the local large scale structure, owing to selection effects, truncation of small scale density fluctuations and uncertainty on galaxy formation processes. Figure~\ref{fig:borg} shows the projected dark matter density in spheres of 50~Mpc for eight different realisations of the constraints with comparable likelihood. Overlaid on each panel are the positions of galaxies in the Local Universe, from the 2MASS survey \citep{Huchra-12}. All eight realisations produce similar large scale features, including the known galaxy clusters and voids in the local universe, but there is still noticeable variation between individual realisations.

\subsection{Converting the {\sc Borg} constraints into an octree function representation}\label{sec:methods:octree_lss}
The large scale structure constraints on the phase information produced
by {\sc Borg}  are expressed as a $256^3$ element Gaussian white noise field.
Taking the Fourier transform of this field we produce an equivalent
octree representation by imposing the  Fourier modes provided by {\sc Borg} as a set of linear
constraints on an unconstrained white noise field generated by
octree functions.

We do this following the method outlined in \cite{Salmon-1996}. This procedure applied to Gaussian white noise fields is particularly simple. In essence, we create an unconstrained Gaussian white noise field in the octree representation, following Section~5.2 of \cite{Jenkins-2013} and compute the Fourier amplitudes of this unconstrained field. We then adjust each of the corresponding Fourier amplitudes to exactly match the corresponding Fourier mode predicted by {\sc Borg}, and then convert back to the octree representation. 

This procedure is not unique as it can be applied to different sets of octree functions depending on the choice of how deep to go within the tree. For this paper, we chose to use the octree information down to level 21. At this level the information content of the octree functions for the simulation volume consists of $240^3$ degrees of freedom, which is a close match to the {\sc Borg} constraints. As the number of degrees of freedom in the octree representation are slightly less than in {\sc Borg},  we miss some of the {\sc Borg} shortest wave Fourier modes as constraints. The {\sc Borg}  constraints that we use are not unique at these scales in any case.  This particular choice means that in practice all the information in the initial unconstrained octree Gaussian white noise field is overwritten and the constraints are effectively confined to definite levels of the octree. This leaves all octree functions at deeper levels of the tree unconstrained, and thus free to be randomly varied.

\subsection{Random Variations} \label{sec:methods:variations}
In a second step, as shown in Figure~\ref{fig:shifts}, we take a sample from {\sc Borg} and supplement it with random phase information at level 22 and above, i.e. randomising phases with $\lc = 3.24$~cMpc and smaller. In total, we create 60~000 variations of the initial density field. We call this set the "3.2~cMpc set", and label individual variants with index $i$. The two top rows of Figure~\ref{fig:spheres} show the projected dark matter density in spheres of radius 5~Mpc centred on the fiducial observers in four "3.2~cMpc" variations. As all variations share the same relevant phase information in the constrained region up to level 21 ($\lc = 6.48$~cMpc), all simulations in this sample have a similar intermediate scale density, including a planar structure and several visible filaments. However, randomising the phase information on levels 22 and above yields different populations of haloes. We describe the properties of the LG analogues found among the "3.2~cMpc" set in Section~\ref{sec:results:firstorder}.

As shown in ~\cite{Sawala-2020}, specifying the phase information up to and including octree level 22 ($\lc = 3.24$~cMpc) determines the existence of individual haloes down to $M_{200} \sim 10^{12} \Ms$, i.e. haloes similar in mass to the MW and M31. After identifying pairs of such haloes among the simulations of the 3.2~cMpc set, we create smaller scale variations of promising candidates by independently randomising the phase information at levels 24 ($\lc = 0.81$~cMpc) and 25 in a small region, as shown on the right of Figure~\ref{fig:shifts}. We refer to these sets as the 0.8~cMpc sets, and label individual variants with an additional index, $j$. The effect of variations at 0.8~cMpc can be seen in the bottom two rows of Figure~\ref{fig:spheres}. Here, a Local Group analogue consisting of a pair of haloes of $\sim 10^{12} \Ms$ is nearly always present, but its properties vary. We describe the results of these simulation in Section~\ref{sec:results-second-order}. Still smaller scale variations, up to level 28 ($\lc = 0.05$~cMpc), are described in Section~\ref{sec:higher-order}.

\subsection{Numerical Setup} \label{sec:methods:simulations}
All simulations were performed with the {\sc Gadget-3} code, a variant of the publicly available code {\sc Gadget-2} \citep{Springel-2005-gadget}. To reduce the computational cost, before running the full set of 60~000 3.2~cMpc variants, we first ran a prototype set of 100, refining a cubic region of $20^3$ cMpc$^3$ around the origin of the fiducial observer within the constraints. Because of the partially identical phase information, the variation in the Lagrangian volumes is limited. In the next step, we computed the union of all particles within $r=8$~Mpc of the fiducial observer at $z=0$ in all 100 variants. This defined the refinement region (or "mask"), ensuring that the region of interest ($r=5$~Mpc) is always included in the high-resolution region, while limiting computational cost. Extracting this region out of a $20^3$~cMpc$^3$ cube sampled with $80^3$ particles resulted in only $\sim 52^3$ particles required in the high resolution region. Each simulation required 4.4 core-hours on the  COSMA-6 computer.

The mass resolution of the 3.2~cMpc variants is $2.4\times 10^9~\Ms$; i.e. a typical MW or M31 halo of $10^{12}~\Ms$ is resolved with $\sim 420$ particles. We used a $480^3$ FFT for computing the initial particle displacements inside the high resolution region. The other variants were performed at twice higher spatial and eight times higher mass resolution, using masks based on the 3.2~cMpc simulations. We tested that the results do not change qualitatively when using higher particle resolution or finer FFTs.

\subsection{Identification and characterisation of LG candidates}\label{sec:methods:identification}
The distance between the MW and M31 has been measured using Hipparcos and HST observations of red clump stars as $d=784^{+13}_{-17}$~kpc \citep{Stanek-1998}, using TRGB stars as $d=785 \pm 25$~kpc \citep{McConnachie-2005}, using Cepheids as $d=765 \pm 28 $~kpc \citep{Riess-2012}. The galactocentric radial velocity has been accurately measured, $v_{\rm r}=-109.3 \pm 4.4$~km~s$^{-1}$ \citep{vanderMarel-2012}. The transverse velocity is more controversial. The only direct measurement of proper motions with HST \citep{vanderMarel-2012} give $v_{\rm t}=17 \pm 17$ km~s$^{-1}$, while \cite{VanderMarel-2019} give a combined estimate from Gaia DR2 and HST data of $v_{\rm t}=57^{+35}_{-31}$ km~s$^{-1}$. Both of these imply a highly radial orbit. Conversely, indirect measurements using satellite kinematics give far higher values: $v_{\rm t}=164.4 \pm 61.8$~km~s$^{-1}$ \citep{Salomon-2016}, and a much lower eccentricity. In this work, we adopt a distance of $d=770$~kpc, a radial velocity of $v_{\rm r}=-$109~km~s$^{-1}$, and a transverse velocity of $v_{\rm t}< 40$~km~s$^{-1}$. Of course, the methods and results described here remain valid for different parameters values.

In our simulations, halos and self-bound subhalos\footnote{In the following we use the term "halo" when referring to a self-bound structure identified by the {\sc Subfind} algorithm \citep{Springel-2001-subfind}, and we consider the mass of the self-bound structure as the mass of the halo. The LG mass is defined as the sum of the two subhalo masses.} were identified in the $z=0$ snapshots using the FoF and {\sc Subfind} algorithms. For identifying Local Group analogues, we consider all possible pairs of self-bound haloes, each in the mass range $5\times 10^{11} - 5\times 10^{12}\;\Ms$, whose centres of potential are separated by $0.5 - 2$~Mpc. We also require that there be no other halo more massive than the smaller of the pair within $2$~Mpc.

In previous works, both those that studied the Local Group in arbitrary environments \citep[e.g.][]{Sawala-2016b} as well as those including constraints \citep{Carlesi-2016}, the orientation of the Local Group with regards to the large scale structure was considered arbitrary. However, in constrained simulations, we argue that reproducing the correct orientation of the Local Group with respect to the observed large scale structure is not a mere accessory. If the distinct features observed in the LG are linked to the local large scale structure, and in particular, if the anisotropy of the LG's environment plays any role in its evolution, then the LG's orientation with respect to this environment would appear to be crucial. And, as we will discuss in Section~\ref{sec:characteristics}, the orientation is also set on fairly large scales.

Our simulation axes are aligned with the equatorial coordinate axes, placing the observed local structures at their observed locations. To parametrise the orientation of the LG analogues relative to the large scale structure, we calculate in turn the vectors pointing from the centre of each halo to the other, and quantify the angles, $\delta_{1,2}$, by which they differ from the true position of M31 relative to the MW. We then designate the haloes as the "MW" and "M31" analogues, respectively, so that $\delta = $min$(\delta_{1,2})$; $\delta$ thus ranges from $0^\circ$ for a LG analogue with the correct orientation to $90^\circ$ for a LG analogue whose orientation is perpendicular to the observed one. Having assigned the MW and M31 counterparts, we parametrise the mass ratio as $\Mr = \mathrm{M_{M31}} / \mathrm{M_{MW}}$. Similarly to the total mass, the mass ratio is also uncertain, with most estimates within a factor of two of unity, and M31 likely to be the more massive of the pair \citep{Li-2008, Phelps-2013, Sofue-2015}.

We also require that the position of the MW be within 5~Mpc of the fiducial observer defined by the cosmological constraints. Subsequently, as outlined in Section~\ref{sec:numbers}, we apply additional limits on the mass, mass ratio, separation, velocities, and orientation of the LG candidates.

\section{Variations at $\lc$~=~3.2~\texorpdfstring{\MakeLowercase c}{c}Mpc} \label{sec:results:firstorder}
A simulation where the density field is constrained down to only $\sim 4$~Mpc does not generally yield a Local Group at its centre. As explained in Section~\ref{sec:methods:variations}, to complement the phase information constrained by the local large scale structure, in a first step, we generated a set of 60~000 simulations, randomly varying the phase information at scales of $\lc \sim 3.2$~cMpc and smaller, that is levels of 22 and above.

\begin{table}
%    \makegapedcells % -- default looks too cramped, imho, but replaced with custom spacings.
    \centering
    \begin{tabular}{l|c|c|c}
         &"Loose" & "Intermediate" & "Strict" \\[1mm]
    \midrule  & & \\[-2mm]
    $\mathrm{M} [10^{12}\Ms] $ & 1.2 \dots 6.0 & 1.5 \dots 5.0 & 2.0 \dots 4.0\\  
     & (46054) & (39090) & (24108) \\[1mm]
    $\Mr$ & 2/5 \dots 5 & 2/3 \dots 3 & 1 \dots 2\\  
     & (40490) & (28630) & (14162) \\[1mm]
    $d$ [Mpc]  & 0.5 \dots 1.5 & 0.6 \dots 1.0 & 0.74 \dots 0.80 \\
     & (34693) & (15356) & (1665) \\[1mm]
    $v_r \mathrm{[kms^{-1}]}$ & -200 \dots 0 & -150 \dots -50 & -109 \dots -99\\
     & (11414) & (4318) & (597) \\[1mm]
    $v_t \mathrm{[kms^{-1}]}$ & < 150 & < 100 & < 40 \\
     & (38885) & (23904) & (10792) \\[1mm]
    \midrule & & \\[-2mm]
    N (M,$\Mr$,$d$,$v_r$,$v_t$)  & 6385 & 489 & 1 \\[1mm]
    \midrule & & \\[-2mm]
    $\delta$ &  $< 45^\circ$ & < $30^\circ$ & < $15^\circ$ \\[1mm]
    N (M,$\Mr$, $d$,$v_r$,$v_t$,$\delta$) & 2309 & 82  & 0 \\[2mm]
    \end{tabular}
    \caption{Selection criteria and numbers of LG analogues among the total of 48,323 candidates from the 3.2~cMpc set. M is the sum of the MW and M31 subhalo masses, $d$ is the distance between the centres of potential, $v_r$ and $v_t$ are the relative velocities, $\delta$ is the error in the position on the sky of M31. Numbers in brackets indicate the number of LG analogues from the 3.2~cMpc variations satisfying the corresponding criterion. N indicates the number that simultaneously satisfy all criteria in each column, either excluding or including the constraint on the orientation.}
    \label{tab:criteria}
\end{table}

\subsection{Number of LG analogues} \label{sec:numbers}
Searching the central $r=5$~Mpc sphere in all 60~000 simulations of the 3.2~cMpc set for Local Group analogues yielded a total of 48,323 halo pairs (hereafter labelled "all") with a Local Group mass in the range $\sim 1-8 \times 10^{12}\;\Ms$, a separation of 0.5 -- 2~Mpc, and fulfilling the condition that no other halo more massive than the lower-mass member be within 2~Mpc of the pair midpoint. As shown in Table~\ref{tab:criteria}, restricting the selection criteria closer to the observed values has the effect of significantly reducing the number of LG analogues: our "loose" criteria for mass, mass ratio, separation, radial velocity and tangential velocity yield 6385 LG analogues, and only one LG analogue fulfils the strict "criteria". Once we also include constraints on the orientation, the numbers are reduced even further, and when we limit the orientation of the LG relative to the LSS to within $15^\circ$, none fulfils all strict criteria.

\begin{figure}
    \includegraphics[width=\columnwidth]{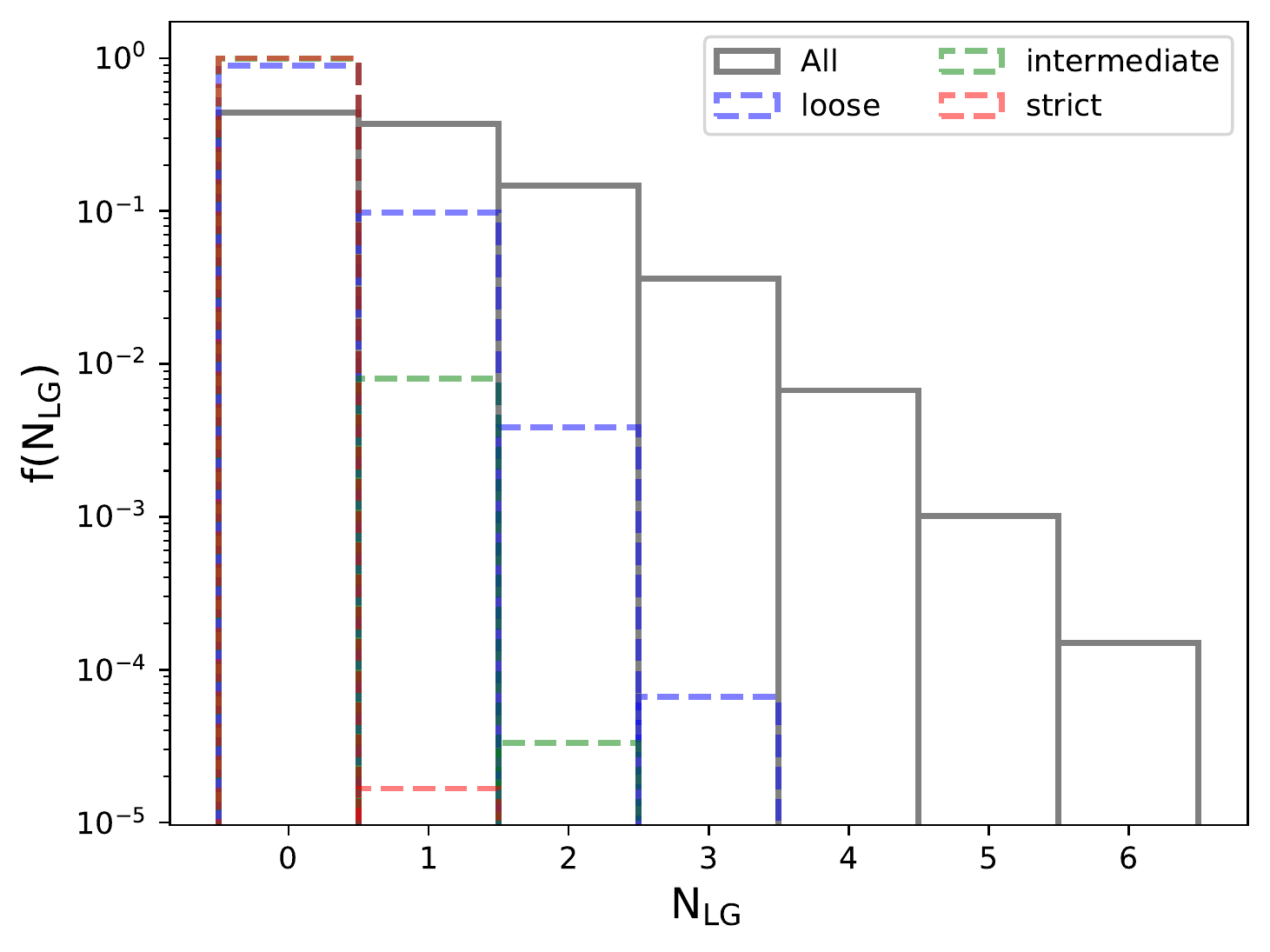} \vspace{-.5cm}
    \caption{Fraction of simulation volumes from the 3.2~cMpc set with $\mathrm{N_{LG}}$ Local Group analogues selected by different criteria. With the "loose" criteria, $9.8\%$ of variants contain one LG analogue, $0.4\%$ contain two, and 4 volumes $(0.0067\%)$ contain three LG pairs. Stricter limits on mass, mass ratio, separation and velocity reduce the numbers, but even more so, the multiplicity: with the intermediate criteria, only 2 variants $(0.0033\%)$ have two LGs and none have more than two. 
    }
    \label{fig:multiplicity}
\end{figure}

\begin{figure}
\begin{overpic}[width=1.63in]{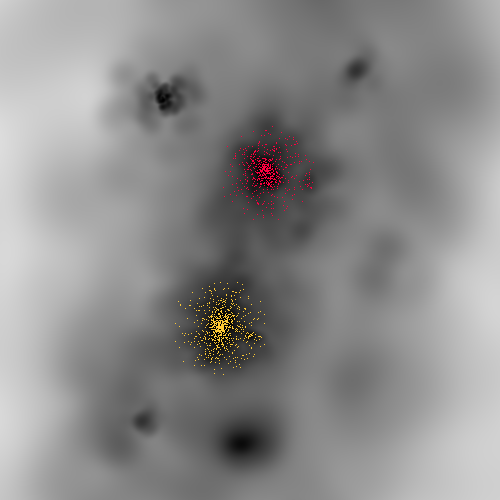} 
 \put (75,90) {\textcolor{white}{\textbf{i=622}}}
 \put (65,65) {\textcolor{white}{\textbf{MW}}}
 \put (55,25) {\textcolor{white}{\textbf{M31}}}
\end{overpic}
\begin{overpic}[width=1.63in]{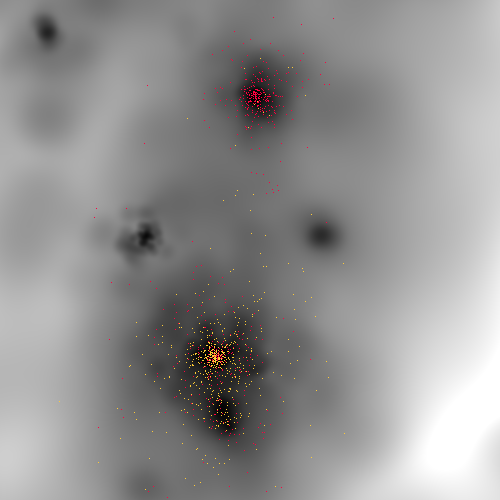} 
 \put (75,90) {\textcolor{white}{\textbf{i=8137}}}
 \put (65,80) {\textcolor{white}{\textbf{MW}}}
 \put (55,25) {\textcolor{white}{\textbf{M31}}}
\end{overpic}\\
\begin{overpic}[width=1.63in]{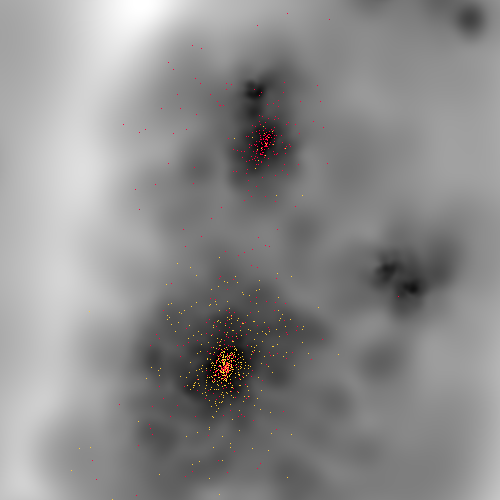} 
 \put (75,90) {\textcolor{white}{\textbf{i=14259}}}
 \put (65,65) {\textcolor{white}{\textbf{MW}}}
 \put (55,25) {\textcolor{white}{\textbf{M31}}}
\end{overpic}
\begin{overpic}[width=1.63in]{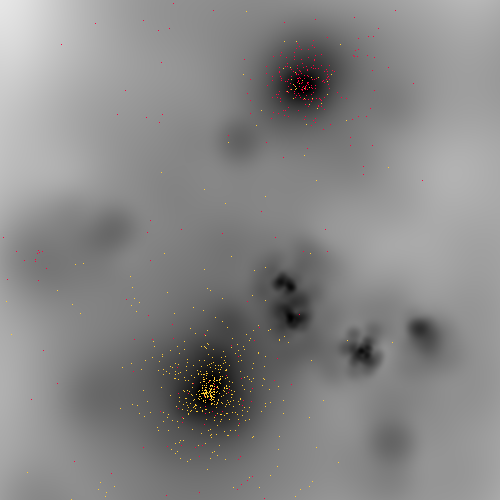} 
 \put (75,90) {\textcolor{white}{\textbf{i=40431}}}
 \put (70,80) {\textcolor{white}{\textbf{MW}}}
 \put (55,20) {\textcolor{white}{\textbf{M31}}}
\end{overpic}
    \caption{Duplicates of a LG analogue from the 3.2~cMpc variations. The top left shows the projected DM density around the LG for $i=622$, highlighting in red and yellow particles within 250 kpc of the centres of the MW and M31 analogue haloes. The other three panels show the dark matter density in three other realisations of the 3.2~cMpc set, at the same $z=0$ coordinates, and with the same particles highlighted. All four LG analogues share more than $50\%$ of particle IDs, i.e. they have formed from the same Lagrangian volume, and also have similar mass and orientation.}
    \label{fig:collisions}
\end{figure}

\begin{figure*}
    \includegraphics[width=7in]{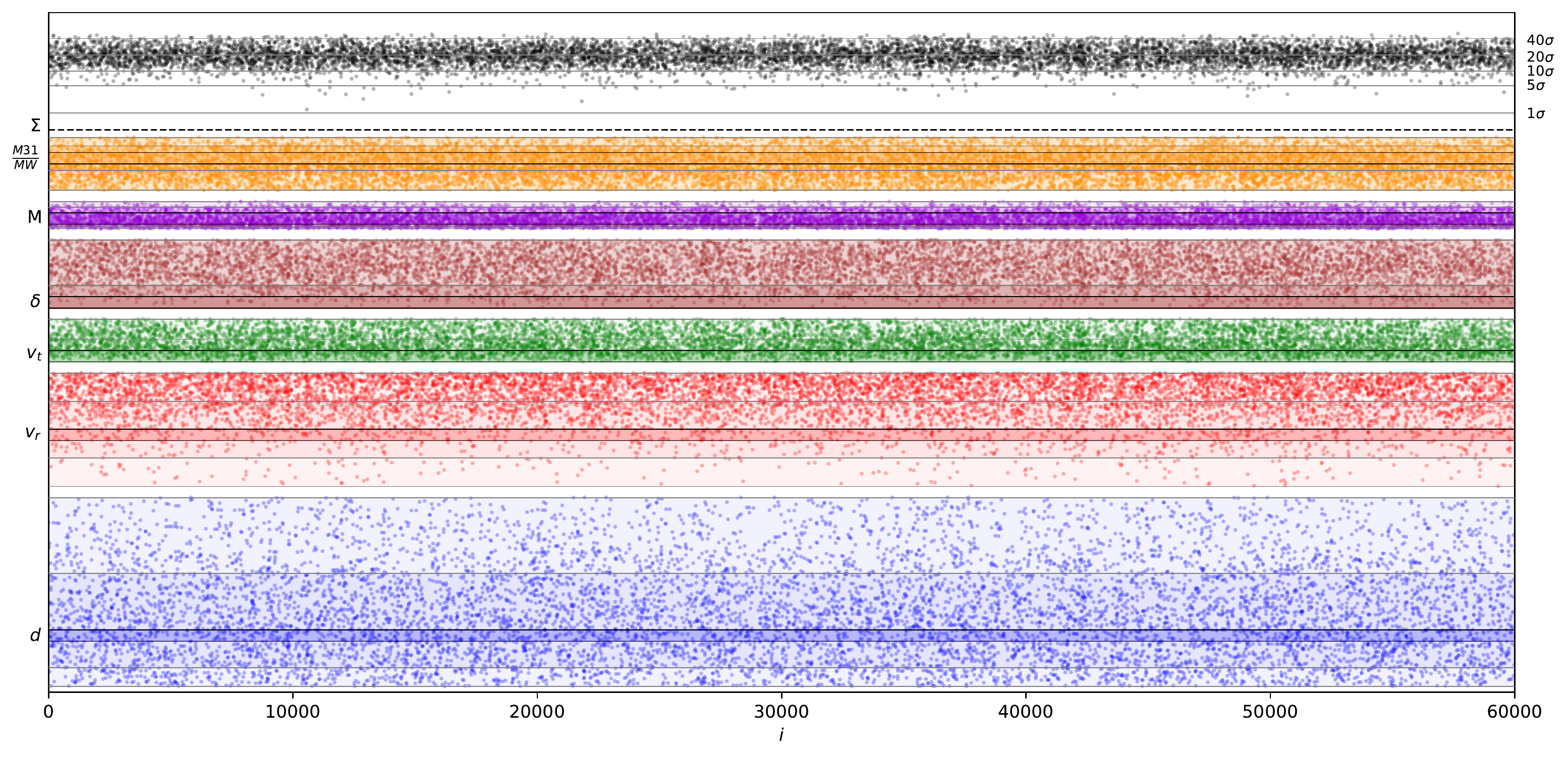} \vspace{-.5cm}
    \caption{Quality of LG analogues for 3.2~cMpc variations. Each vertical set of points represents a LG analogue from a simulation with index, $i$. Sets of horizontal stripes indicate the logarithm of the mass ratio ($\Mr$, orange), the mass ($M$, purple), angular offset ($\delta$, brown), tangential velocity ($v_t$, green), radial velocity ($v_r$, red) and separation ($d$, blue), respectively. For each observable, dashed lines indicate either the $1\sigma$ observational uncertainty, or the accepted error for our "strict" criteria. Grey points on top show $\Sigma$, the total error outside the accepted range for all observables. While many variants fulfil the criteria for individual observables and some fulfil multiple criteria, no points fall on the thick dashed line, i.e. none of the 3.2~cMpc variants matches all six criteria simultaneously.
    }
    \label{fig:quality-first-order}
\end{figure*}

\subsection{Multiplicity} \label{sec:multiplicity}
With the broadest selection criteria, labelled "all", a single simulation often contains more than one LG analogue within the central $r=5$~Mpc volume. Figure~\ref{fig:multiplicity} shows the number of variants from the 3.2~cMpc set containing $N$ Local Group analogues for different selection criteria. As the selection criteria are tightened, the number of variants that contain multiple Local Group analogues decreases sharply, and more steeply than $N^{-2}$. With the "loose" constraints, 9.8\% of variants have one LG, but only 0.4\% contain two and none contain more than three. With intermediate criteria, while there are 489 LGs, the chance of finding two in the same volume is $\sim 1:30~000$.

More interestingly, we not only find multiple different LG analogues within the same $r=5$~Mpc volume, we also find many instances of ostensibly the {\it same} LG forming in multiple volumes among the 3.2~cMpc variations. For the purpose of finding such duplicates, we consider two LGs in different simulations to be the same object if their centres of mass differ by less than 300 kpc, their total masses differ by less than $30\%$, and their orientations are aligned to within 30$^\circ$. We find that only $\sim 10\%$ of LGs from either the loose or intermediate set have {\it no} duplicates, while almost $\sim 30 \%$ of LGs are found more than 10 times.

In Figure~\ref{fig:collisions}, we show four "copies" of a LG analogue from the 3.2~cMpc set. All four panels are centred on the same coordinates, and all four contain a pair of haloes with similar orientation. As described in Section~\ref{sec:methods:simulations}, all 3.2~cMpc variants use the same particle load, so that particles with the same IDs have identical (unperturbed) Lagrangian coordinates in all simulations from this set. Highlighted in red and yellow on all four panels are the positions of particles within 250~kpc of the centres of the MW and M31 analogues shown in the top-left panel. In each of the other three simulations, a large fraction of the same particles are also found in the two haloes. In general, we find that copies of LGs share a large fraction of particles, indicating that they originate from largely overlapping Lagrangian volumes.

The fact that most of our LG analogues exist more than once suggest that, within the 60~000 random realisations of the 3.2~cMpc set, we are likely to have exhausted a large fraction of the possible, truly distinct LG analogues that can form given this particular set of constraints. As we will discuss in the conclusions (Section~\ref{sec:conclusion}), this has important implications for our future ability to test cosmological models. In the following sections, we make use of the fact that we can generate several variations of the same LG by deliberately introducing small-scale random variations to promising LG candidates identified among the 3.2~cMpc variants.

\subsection{Quality of LG analogues}\label{sec:quality-first-order}

Figure~\ref{fig:quality-first-order} illustrates the quality of individual LG analogues for all 8086 3.2~cMpc variations that satisfy the "loose" criteria irrespective of orientation, as given in Table~\ref{tab:criteria}. On the x-axis, we show the index, $i$, while on the y-axis, we quantify the agreement with the observation in terms of the logarithm of the mass ratio mass ($\Mr$, orange), the total mass (M, purple), orientation ($\delta$, brown), tangential velocity ($v_t$, green), radial velocity ($v_r$, red) and separation ($d$, blue). For each variable, the set of solid black lines correspond to our strict criteria (see Table~\ref{tab:criteria}), either the observational uncertainty, or in the case of the orientation, where the observational uncertainty is negligible, an acceptable error of $\delta=15^\circ$. The grey lines denote the intermediate and loose criteria.

The vertical positions of individual points of corresponding colour are normalised relative to the strict criteria, and show 
\begin{equation}
\Delta=\frac{x_{n,i} - x_n}{\sigma_n},
\end{equation}
where $x_{n,i}$ is the value of variable $x_n$ for simulation $i$, $x_n$ is its observational value, and ${\sigma_n}$ is the observational uncertainty or acceptance region. For clarity, different variables are offset vertically. The grey points at the top of the figure show the sum of the individual errors that lie outside the observational bounds, i.e.
\begin{equation}
\Sigma_i = \sum\limits_{n} \left( \mathrm{min} \left( 1, \frac{|x_{n,i} - x_n|}{\sigma_n} \right) \right) - n.
\end{equation}
A LG candidate that lies within the bounds for all five variables would have a value of $0$ and fall on the thick dashed line on the plot. The fact that all grey points lie above this line indicates that $\Sigma > 0$ for all candidates, i.e. all candidates from the 3.2~cMpc set are outside the bounds for at least one observable.

\section{Variations at $\lc$~=~0.8~\texorpdfstring{\MakeLowercase{c}}{c}Mpc} \label{sec:results-second-order}
As discussed in the previous section, our set of 60~000 random variations produced tens of thousands of LG analogues within the constraints, and indeed, resulted in many variants of ostensibly the {\it same} objects, but it did not yield any that matched the strict criteria for the MW-M31 pair, and only two which match the strict constraints neglecting the orientation. While it would be possible simply to create more tries, a back-of-the-envelope estimate suggests that it would require more than a million simulations in order to find a realisation that satisfies all constraints simultaneously.

Instead, we select the most promising candidates for which a change of smaller scales in the initial density field might produce variations that fulfil the constraints. For these, we keep the phase information fixed for two further levels above the constraints, 22 and 23, and independently vary levels 24 and above (corresponding to scales with $\lc <$ 0.8~cMpc). As discussed in Section~\ref{sec:methods:variations} we expect that a majority of these variants will contain the {\it same} individual $\sim10^{12}~\Ms$ haloes, but with a scatter in their masses, positions and velocities.

\begin{figure*}
    \includegraphics[width=7.2in]{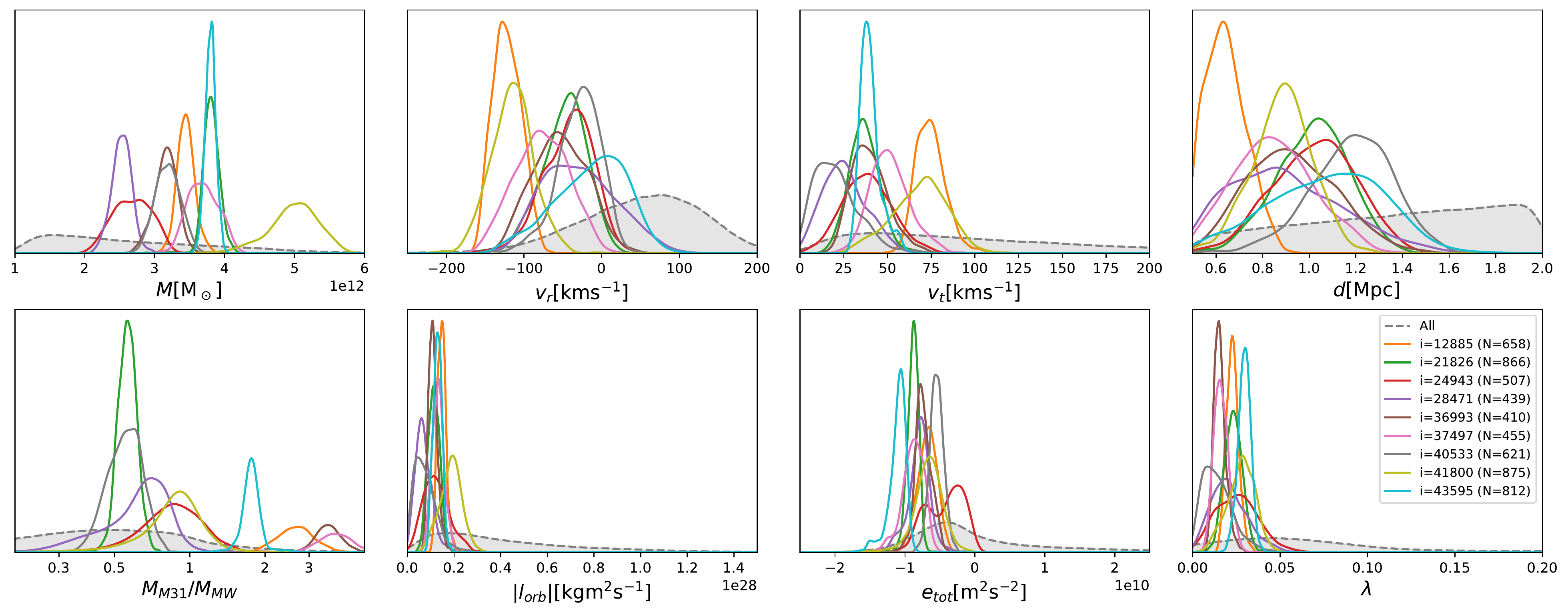}\vspace{-3mm}
    \caption{Probability density functions of mass, radial velocity, transverse velocity and separation (top row) and mass ratio, specific orbital angular momentum, specific orbital energy, and orbital spin parameter (bottom row), for the 3.2~cMpc variants (grey dashed lines), and for the nine sets of 0.8~cMpc variations, independently normalised and coloured according to the index, $i$. The quantity, N, gives the number of LG analogues matching the "loose" constraints. For each set of 0.8~cMpc variants, the scatter is smaller than in the 3.2~cMpc variants in all observables, but while some observables (such as radial velocity and distance) still have comparable scatter, for other variables (e.g. mass, energy and orbital angular momentum), the scatter is greatly reduced and depends on the set.}
    \label{fig:stats:1D}
\end{figure*}

\begin{figure}
    \includegraphics[width=\columnwidth]{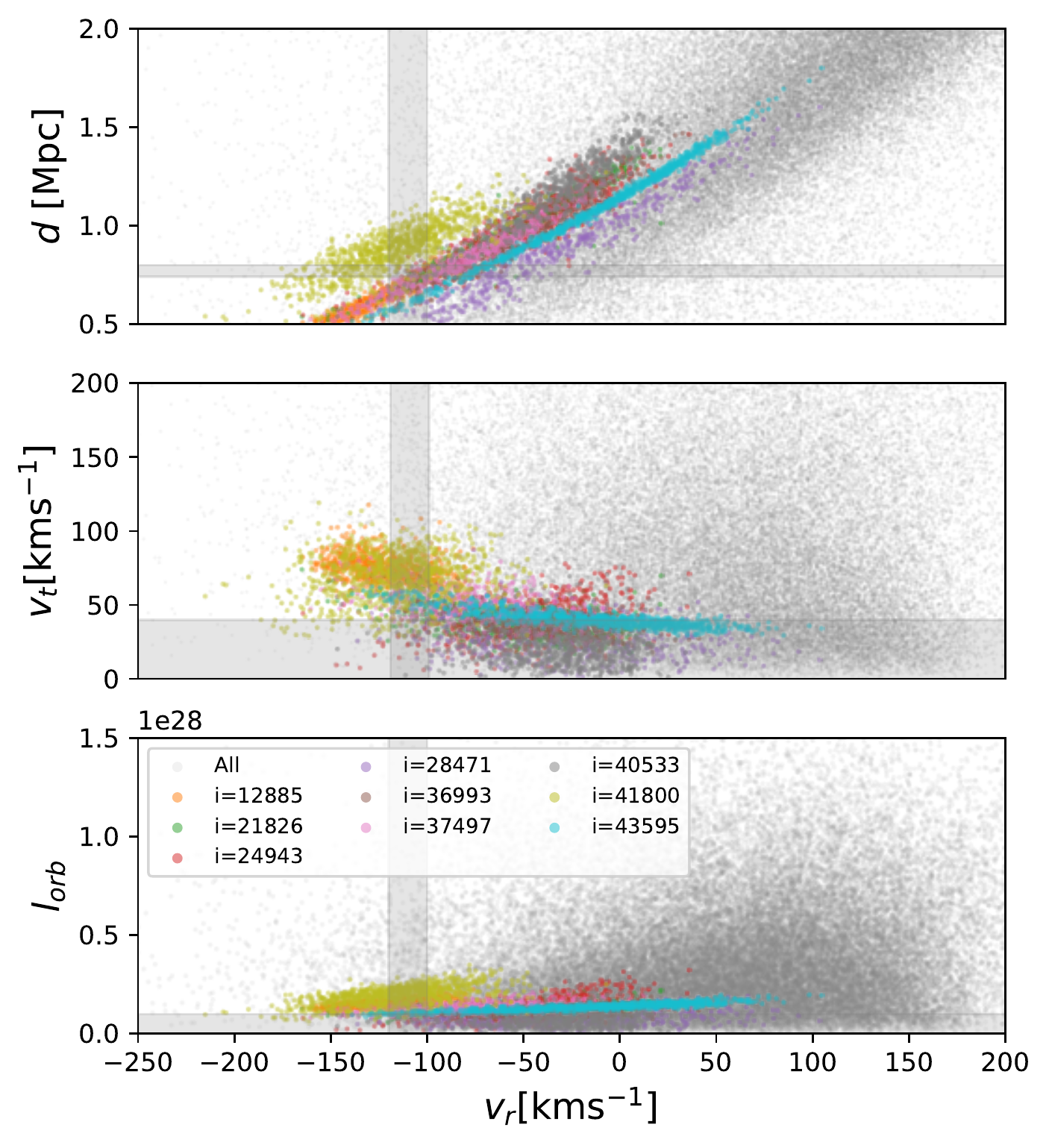} \vspace{-3mm}
    \caption{Radial velocity vs separation (top panel), transverse velocity (middle panel), and specific orbital angular momentum (bottom panel), for LG analogues in the 3.2~cMpc sample (grey points), and in the nine sets of 0.8~cMpc variations, coloured according to $i$. Grey bands show our "strict" criteria for $d$, $v_r$, $v_t$, and the corresponding specific orbital angular momentum, $l_{orb}$. There is still considerable scatter in both $v_r$ and $d$ among the variants of each the 0.8~cMpc set, but they only occupy a narrow region of phase space, corresponding to LG orbits of similar energy but different orbital phase. 
    } \vspace{-3mm}
    \label{fig:stats:D-vr-first-second}
\end{figure}

\subsection{Characteristics of the Local Group} \label{sec:characteristics}
To select suitable candidates for refinement, it is important to determine first which properties of the LG are set on large scales in the primordial density field, and which ones vary when smaller scales are modified. As discussed in Section~\ref{sec:methods:identification}, observationally, the Local Group is characterised primarily and most precisely by the separation of the Milky Way and Andromeda, their radial velocity, their transverse velocity, and the Local Group's orientation relative to the large scale structure. It may seem natural to select those candidates among the 3.2~cMpc variations that come closest to matching all of these criteria. However, as shown in Figure~\ref{fig:stats:1D}, we find that when we introduce variations on smaller scales, some of these observables still show a great amount of scatter, while others appear to be already set on the scales that determined the {\it existence} of the MW and M31 haloes themselves. In total, we selected nine LG variants among the 3.2~cMpc sets, and created 1000 further variations of each, randomising only phase information at scales of $\lc = 0.81$~cMpc and smaller. 

\begin{figure}
    \includegraphics[width=\columnwidth]{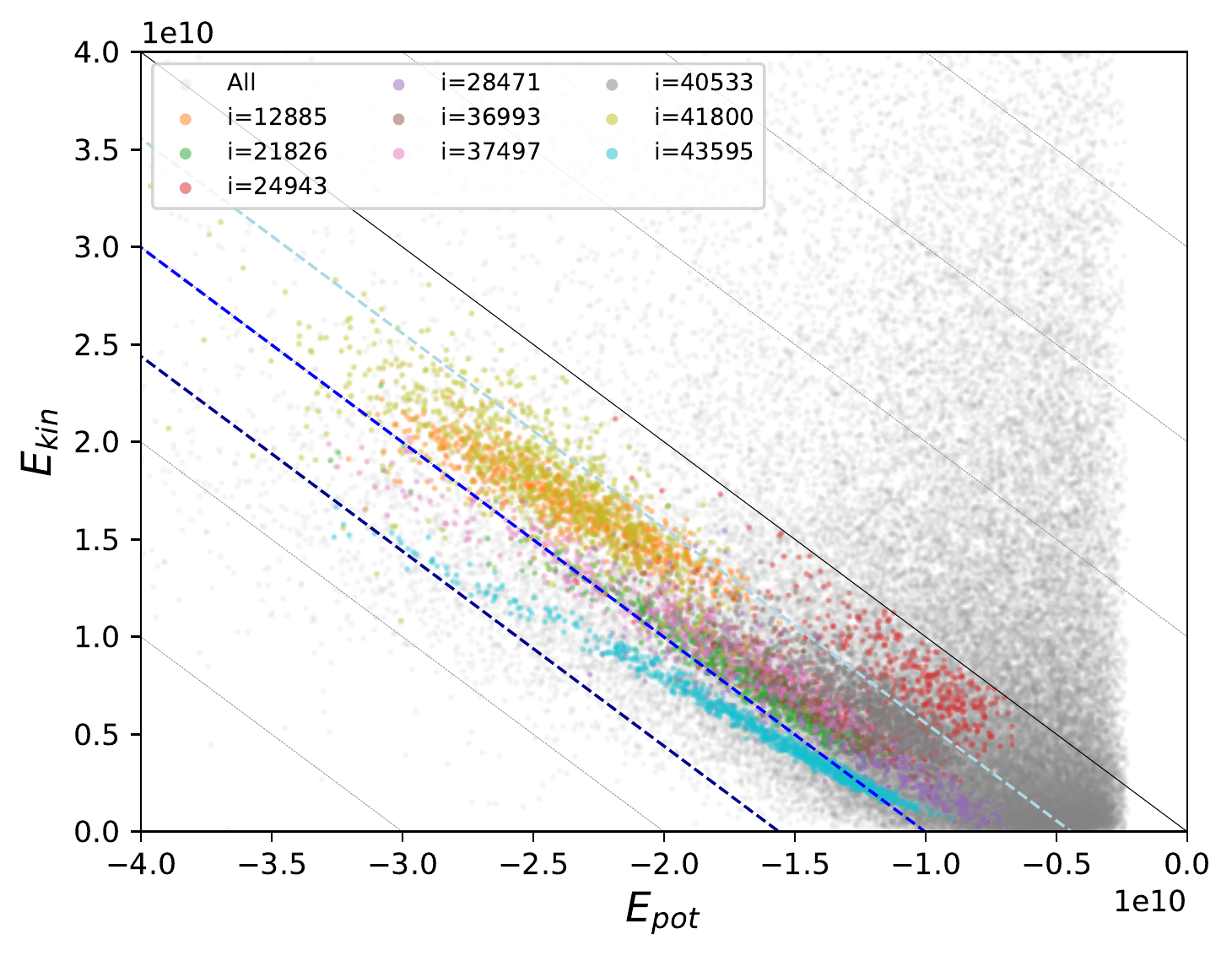}\vspace{-3mm}
    \caption{Specific potential and specific kinetic energy of all LG analogues in the 3.2~cMpc sample (grey dots), and of the LG analogues from the 0.8~cMpc variations, coloured according to $i$. Diagonal lines show constant total energy, the black line marks zero energy orbits. Light, medium and dark blue lines show the energy for LG analogues with $d=770$~kpc, $v_r=-109$~kms$^{-1}$, $v_t=40$~kms$^{-1}$, and with total masses of $2\times10^{12}$, $3\times10^{12}$ and $4\times10^{12}\Ms$, respectively. LGs in the same set appear to have similar total energy, although there is a slight trend for closer pairs to have more negative total energy.}
    \label{fig:stats:epot-ekin}
\end{figure}

\begin{figure}
    \includegraphics[width=\columnwidth]{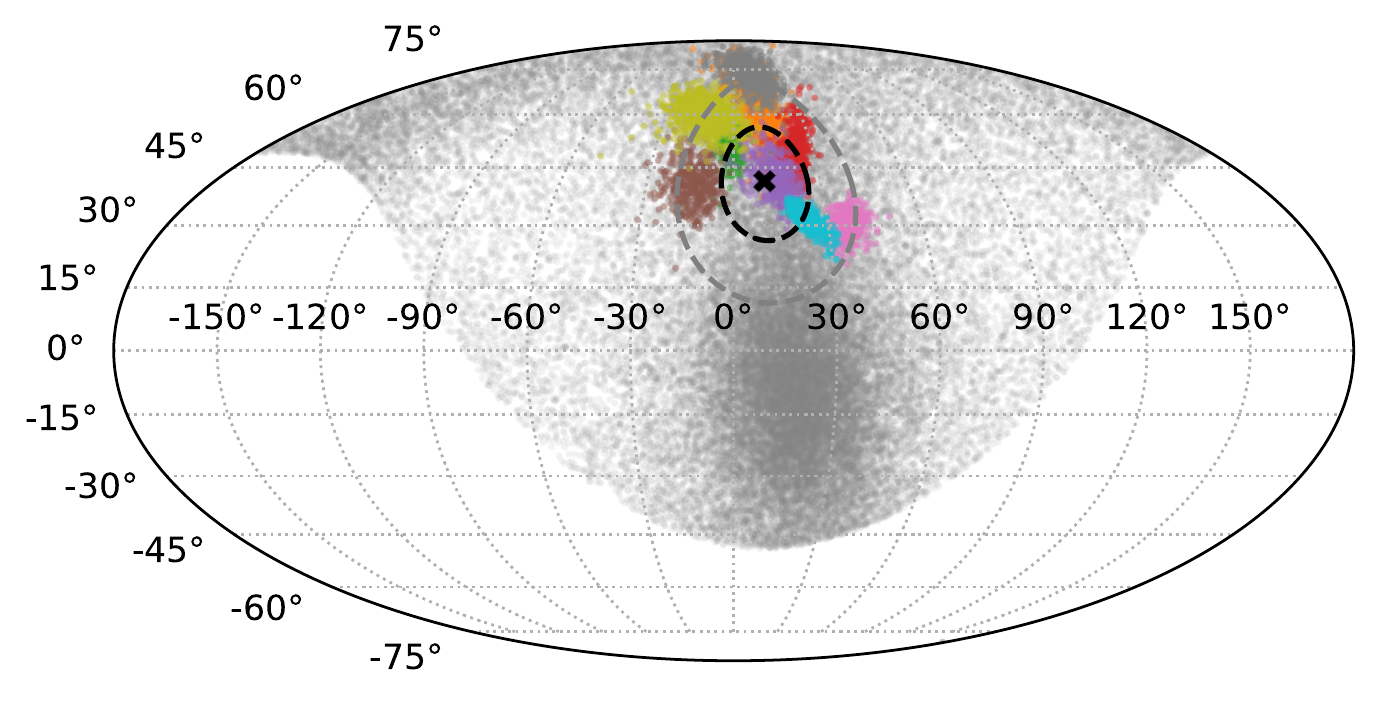} 
    \caption{Mollweide projection of the position of M31 on the sky, for all LG analogues in the 3.2~cMpc sample (grey dots), and for the LG analogues from the 0.8~cMpc variations, coloured according to index $i$ as in Figures~\ref{fig:stats:D-vr-first-second} and \ref{fig:stats:epot-ekin}. The black and grey dashed circles indicate $15^{\circ}$ and $30^{\circ}$ degrees around the true position of M31 observed from the MW, marked by the black {\bf x}. Because we assign the "MW" and "M31" analogues by orientation, the maximum difference from the true orientation is $90^\circ$. Points from the 0.8~cMpc variations cluster closely together and close to the true orientation.
    }
    \label{fig:stats:mollweide}
\end{figure}

The top panel of Figure~\ref{fig:stats:D-vr-first-second} shows the correlation of radial velocity and separation for the 3.2~cMpc variants (in grey), and for the nine sets of 0.8~cMpc variants, coloured according to index $i$. As expected, not all combinations are equally likely, even for the 3.2~cMpc set: more distant LG pairs are typically receding from one another, closer pairs are typically approaching. However, the scatter in the individual 0.8~cMpc sets is remarkably reduced: while distances and radial velocities independently still show a large scatter among the variants in each set, their combinations only span a narrow region. As we discuss below, this is evidence that the 0.8~cMpc variations result in LGs of similar orbits, but differing orbital phases.

The centre panel in Figure~\ref{fig:stats:D-vr-first-second} shows the radial velocity and the magnitude of the transverse velocity. Compared to halo pairs with the same mass, separation, and radial velocity, the small transverse velocity of the Local Group is unusual \citep{Li-2008, Fattahi-2015}. We find that for each set of 0.8~cMpc variations, the scatter in transverse velocity is narrower than the scatter in radial velocity, and this ratio is greater for the LG analogues on more radial orbits. Again, this is expected if each set of 0.8~cMpc variants produces LG analogues of similar, mostly radial orbits whose eccentricities do not vary much.

\subsubsection{Integrals of motion}\label{sec:constants}
The distribution of distances and velocities along orbital trajectories suggests that we should consider as more fundamental the quantities that remain constant in orbital motion. If we consider the LG as a two body system comprised of the MW and M31, these are the total mass, $M = M_\mathrm{MW} + M_\mathrm{M31}$, the specific orbital energy,
\begin{equation}
    e_\mathrm{tot}= \frac{1}{2} \vect{v}^2 - \frac{G M}{r},
\end{equation}
the specific orbital angular momentum,
\begin{equation}
\vect{l}_\mathrm{orb} = \vect{r} \times \vect{v},
\end{equation}
and the reduced Laplace-Runge-Lenz vector, which defines the eccentricity and the orientation of the orbit,
\begin{equation}
\vect{a}_\mathrm{LRL} = \frac{\vect{v} \times \vect{l_\mathrm{orb}}}{G M} - \hat{\vect{r}},
\end{equation}
where $\vect{v}$ is the velocity of M31 relative to the MW, $\vect{r}$ is the position of M31 relative to the MW, $r = |\vect{r|}$ is the distance between the two, and $\hat{\vect{r}} = \vect{r} / r$ is the unit vector from the MW to M31.

From Figure~\ref{fig:stats:1D} and the bottom panel of Figure~\ref{fig:stats:D-vr-first-second}, it can be seen that the 0.8~cMpc variants show little scatter in specific angular momentum compared to the 3.2~cMpc variants. Figure~\ref{fig:stats:epot-ekin} shows the distributions of the 3.2~cMpc and 0.8~cMpc variants in the kinetic and potential energy plane. The LG analogues from a particular set, while greatly differing in both kinetic and potential energy, align along lines of similar total energy. This suggests that variances in radial velocity and separation reflect, to a large extent, a variance in orbital phase.

In Figure~\ref{fig:stats:mollweide}, we show the positions on the sky of M31 in all LG analogues from the 3.2~cMpc sample (in grey), and in the nine sets of 0.8~cMpc variations, in a Mollweide projection. The assignment of the MW and M31 analogues by the LG's orientation limits the position of the M31 analogue to $90^\circ$ around the true position of M31 (indicated by the black cross). Due to the common constraints, the orientations among the 3.2~cMpc set is not uniform; some orientations are more likely than others. In each set of 0.8~cMpc variants, the scatter is significantly reduced. This indicates that the orientation of the orbital plane, and $\vect{a}_\mathrm{LRL}$, are also conserved along the orbit.

For all 0.8~cMpc variants, the orientation close to the observed value, and all orientations within a single set of 0.8~cMpc variants are similar to one another. This result also echoes the findings from the multiplicity of LG analogues among the 3.2~cMpc variations described in Section~\ref{sec:multiplicity}: when we find multiple pairs of haloes of similar mass and in a similar position, their orientations also tend to be similar.

\begin{figure}
    \includegraphics[width=\columnwidth]{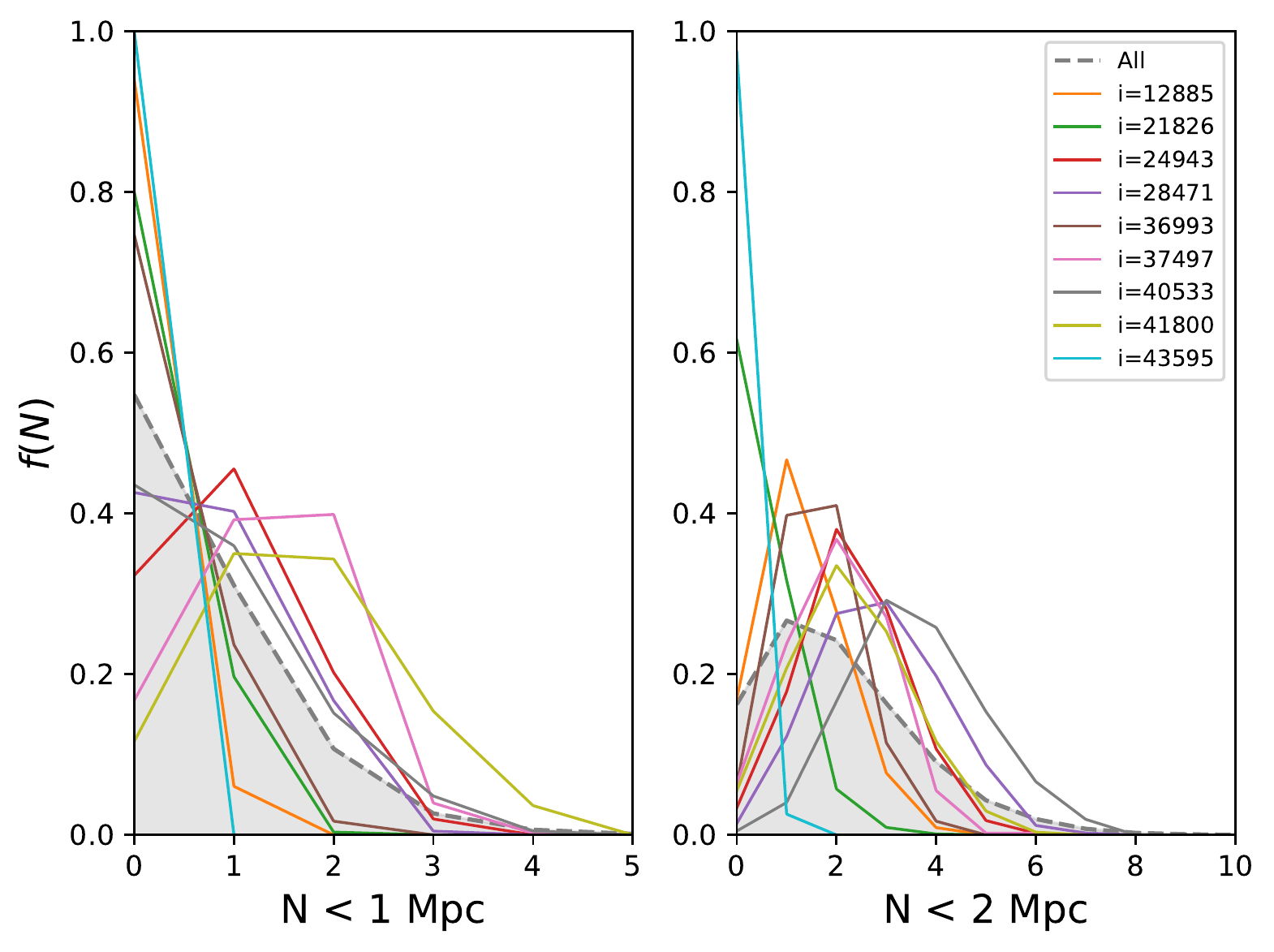}
    \caption{Distribution of the number of objects in the mass range $1-3 \times 10^{11}\Ms$ within 1 or 2 Mpc from the centre of the LG analogue, for the set of 3.2~cMpc variations (grey area), and for the nine sets of 0.8~cMpc variants (coloured lines). The average number of massive satellites in the 0.8~cMpc sets varies, but the scatter within the each set sets is smaller than that of the 3.2~cMpc set.}
    \label{fig:third-2mpc-5mpc}
\end{figure}

It seems clear that these constants of motion are set on larger scales than the canonical observables such as separation and velocities, and while they are more complicated to observe, it seems more natural to consider them as the defining properties of the Local Group. In Section~\ref{sec:higher-order}, we will consider still smaller scale variations to determine the scales that set the phase of the orbit, but as we will see in the next section, the degree to which the Local Group is a true two-body system also affects the scatter in these integrals of motion.

\begin{figure*}
    \includegraphics[width=7.2in]{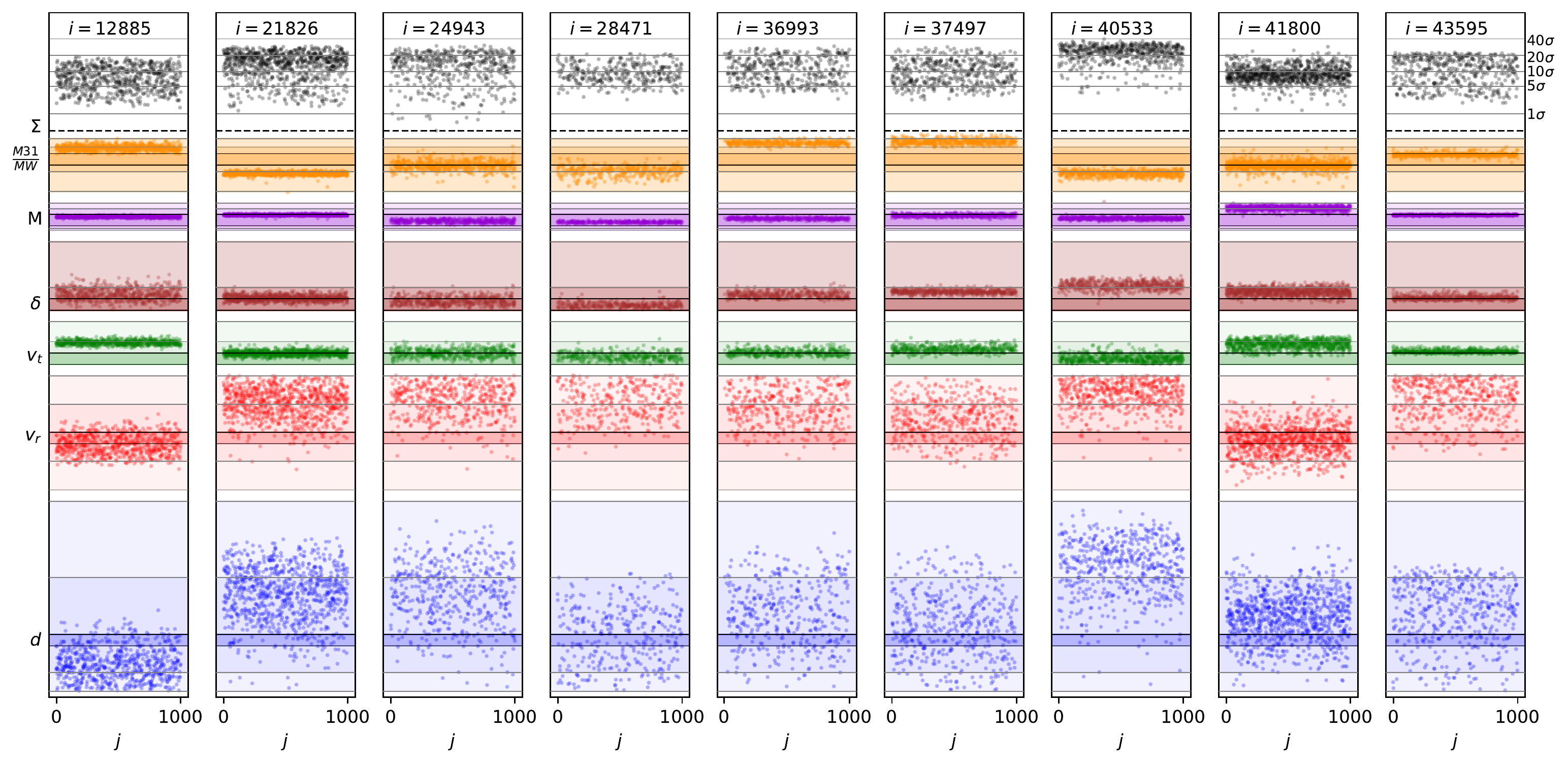}
    \caption{Quality plot of 0.8~cMpc variations, $j$, of 3.2~cMpc variations, $i$, analogous to Figure~\ref{fig:quality-first-order}. There is very little scatter in total mass, angle or, in most cases, mass ratio, less scatter for $v_t$, but still a great amount of scatter in $v_r$ and $d$. The total error, $\Sigma$, is significantly reduced, and the remaining error is now dominated by $v_r$ and $d$. In some cases, the (largely conserved) energy of the orbit (e.g. $i=43595$) or angular momentum (e.g. $i=12885$) imply that no correct combination of $d$ and $v_r$ or correct value of $v_t$ is possible. However, the set with $i=24943$ contains one LG analogue ($j=366$) that matches all "strict" constraints.
    }
    \label{fig:quality-second-order}
\end{figure*}

\subsection{Presence of massive third objects} \label{sec:third-obects}
In addition to M31 and the Milky Way, the Local Group contains two galaxies believed to be residing in haloes of at least $10^{11}\;\Ms$: M33, at a distance of $203\pm37$ kpc from M31 \citep{vanderMarel-2012}, and the LMC at a distance of $\sim50$~kpc \citep{vanderMarel-2002} from the MW centre. With stellar masses of $3-5 \times 10^9\;\Ms$, abundance matching places both in dark matter haloes of $\sim 10^{11}\;\Ms$. HI kinematics further suggests that M33 could have a total mass as high as $5\times 10^{11}\;\Ms$ if it resides in an NFW-like DM halo \citep{Kam-2017}. For the LMC mass, timing constraints give a total mass of $2.5 \times 10^{11}\Ms$ \citep{Penarrubia-2015}; analogues by stellar mass in the {\sc Eagle} simulations indicate $\mathrm{M}_{200} = 3.4^{+1.8}_{-1.2} \times 10^{11}\; \Ms$ \citep{Shao-2018}, while its effect on the Orphan stream in the Milky Way suggests a mass of $1.38^{+0.27}_{-0.24} \times 10^{11} \Ms$ \citep{Erkal-2019}. 

In Figure~\ref{fig:third-2mpc-5mpc}, we show histograms of the number of M33 or LMC analogues, which we define to be in the mass range $1-3 \times 10^{11}\;\Ms$ within 1 and 2 Mpc of the LG centre, for the LGs from 3.2~cMpc variants, and for each of the nine sets of 0.8~cMpc variants.

While the nine sets of 0.8~cMpc variants collectively span a large range in the number of LMC/M33 analogues, we find that some sets ($i=12885, 21826, 36993, 43595$) have no LMC/M33 counterparts within 1~Mpc in a large majority of cases, while other sets ($i=24943, 37497, 41800$) typically have one or more. To a large extent, the presence of such haloes is already determined by the scales shared within each set.

As can be inferred from Figure~\ref{fig:stats:D-vr-first-second} and Figure~\ref{fig:stats:epot-ekin}, the number of massive additional haloes appears to correlate with the scatter in the integrals of motion of a two-body orbit, as discussed in Section~\ref{sec:constants}. For example, the sets of 0.8~cMpc variants with $i=21826$ and 43595 (shown by green and cyan lines or symbols), which typically have no LMC/M33 analogues within 1~Mpc, show very little scatter in total energy, mass or angular momentum, while sets with a higher number of LMC/M33 analogues, such as $i=37497$ and $41800$, (pink and olive lines and symbols) have a much greater scatter in the same quantities.

\subsection{A Local Group from 0.8~cMpc variations}
Figure~\ref{fig:quality-second-order} shows the quality of the LG  in the nine sets of 0.8~cMpc variants, analogous to Figure~\ref{fig:quality-first-order} discussed in Section~\ref{sec:quality-first-order}. We find that the typical mass error, and the error in transverse velocity, is much reduced compared to the 3.2~cMpc variations shown in Figure~\ref{fig:quality-first-order}. The error in orientation clusters close to 0$^\circ$. Reflecting the discussion in Section~\ref{sec:characteristics}, we find that the  scatter in distance or radial velocity, independently, are not substantially smaller than in the 3.2~cMpc set. However, because of the total energy constraint, $d$ and $v_r$ are not independent for LG analogues in the 0.8~cMpc sets. For some sets, this means that no LG analogue matches the observations of both simultaneously, while for others, it significantly increases the fraction of LG analogues which match both observables simultaneously.

If it is possible for a Local Group to form inside the large scale structure constraints, independently randomising all levels simultaneously would eventually yield a Local Group (the 0.8~cMpc variants are, strictly speaking, a proper subset of a (much larger) set of possible 3.2~cMpc variants). The 0.8~cMpc variants bifurcate: because the scatter in the observables is much reduced, for almost all possible sets of 0.8~cMpc variations, it would be impossible to form a LG that matches the observations. Even among the set of nine promising 3.2~cMpc variants we selected, we find examples where 0.8~cMpc modifications will never lead to a formation of a LG, either because the transverse velocity is always too great ($i=12885$), or because the total energy is never negative enough ($i=21826$ and $i=43595$).

However, set $i=24943$ contains one LG analogue [i=24943, j=366] that matches all observational constraints: $\mathrm{M_{LG}} = 2.8 \times 10^{12}~\Ms$, $\Mr=1.07$, $d=769$ kpc, $v_r = -100$~kms$^{-1}$, $v_t = 37$~kms$^{-1}$, $\delta=9.4^\circ$. As discussed in Section~\ref{sec:third-obects}, similar to $i=28471$ and $i = 37497$, variants from this set typically contain several LMC/M33 mass objects, and the scatter in the transverse velocity and the total energy of the two-body orbit is larger than for sets that contain none.

This does not necessarily imply that the presence of LMC/M33 analogues is required to form a LG with the observed parameters (although it is necessary in our particular LSS constraints), but we can conclude that the presence of the LMC and M33 significantly affects the kinematics of the MW-M31 orbit.

\begin{figure}
\begin{overpic}[width=.81in]{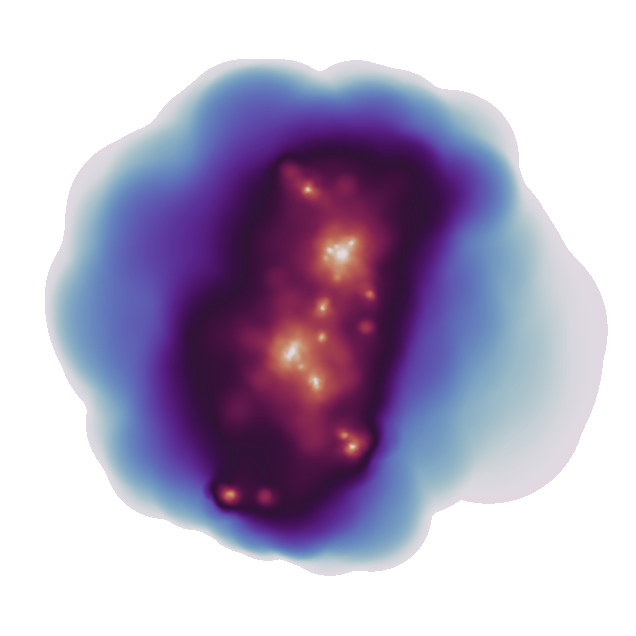}
\put (0,90) {\textcolor{black}{\textbf{L=25}}}
\end{overpic}
    \includegraphics[width=.81in]{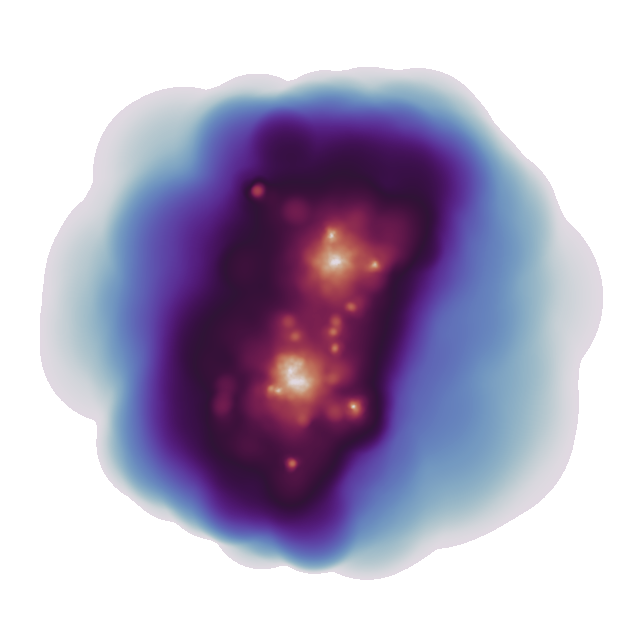}
    \includegraphics[width=.81in]{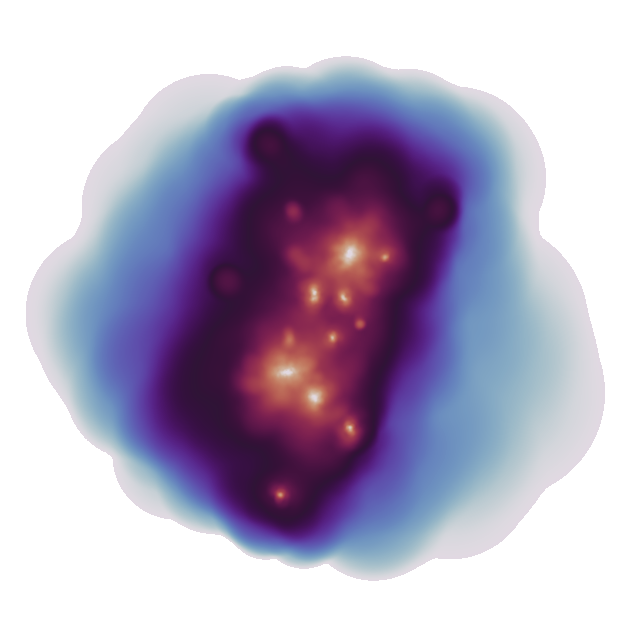}
    \includegraphics[width=.81in]{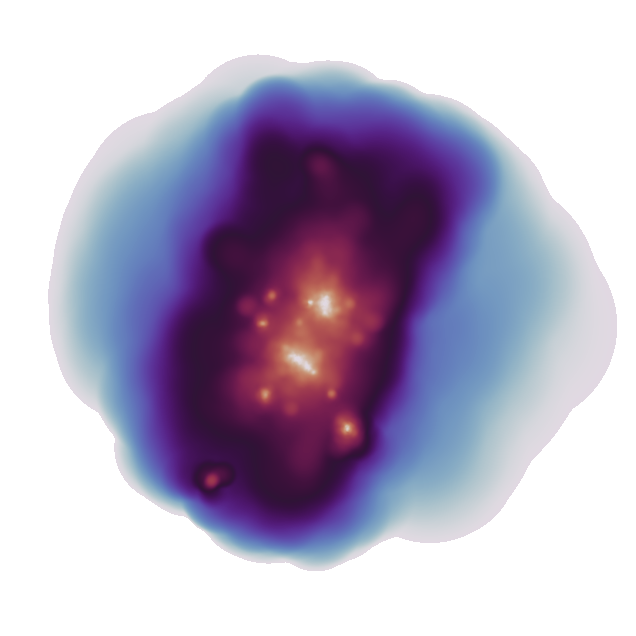}
\begin{overpic}[width=.81in]{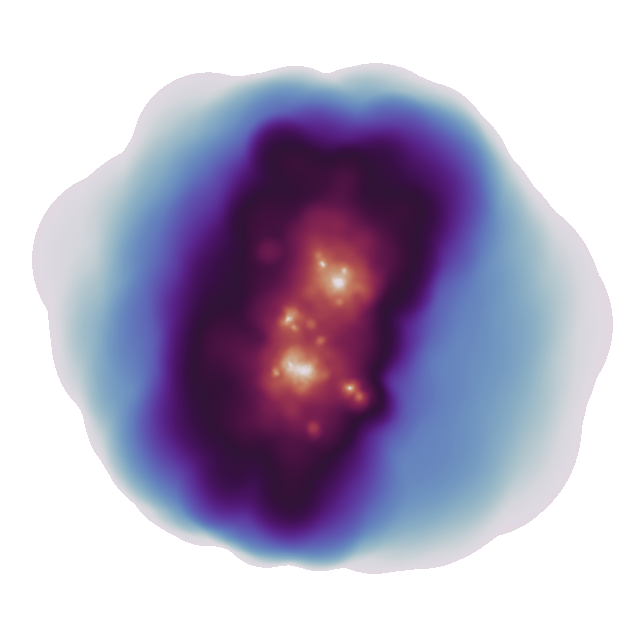}
\put (0,90) {\textcolor{black}{\textbf{L=26}}}
\end{overpic}
    \includegraphics[width=.81in]{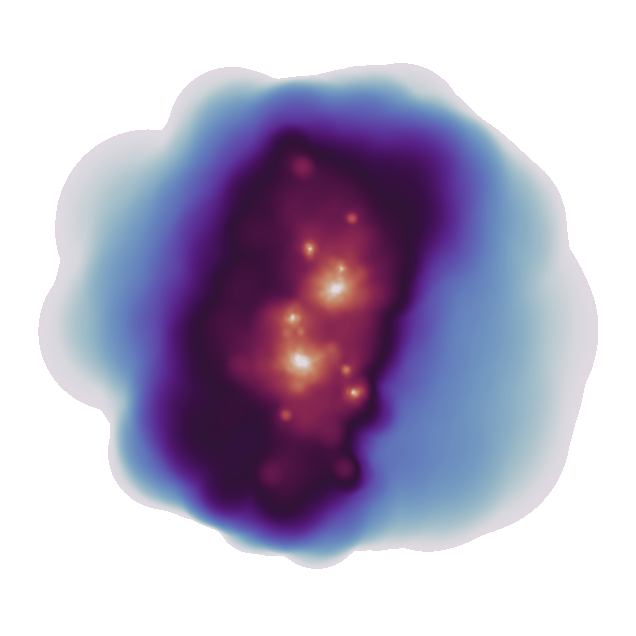}
    \includegraphics[width=.81in]{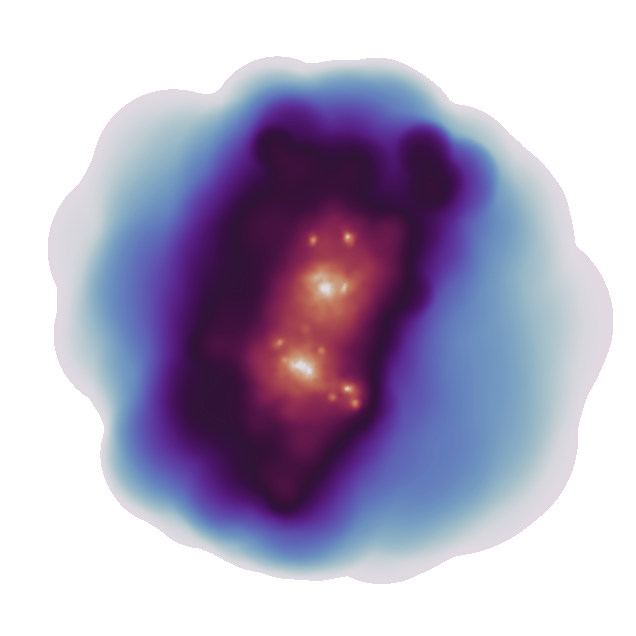}
    \includegraphics[width=.81in]{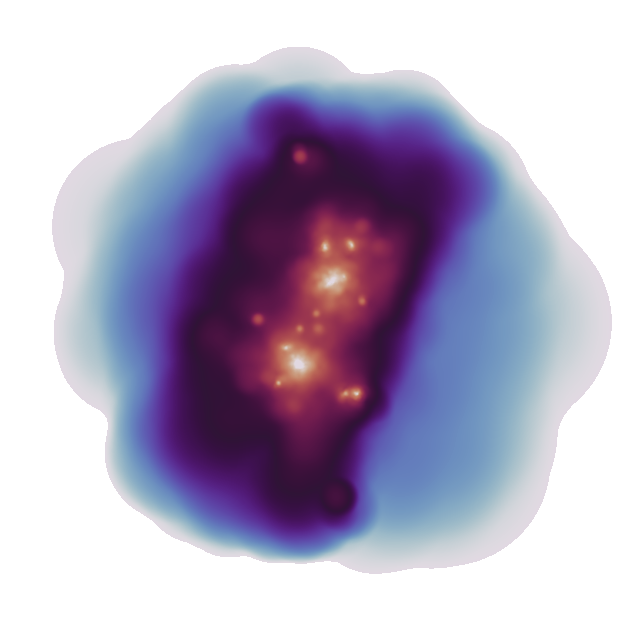}
\begin{overpic}[width=.81in]{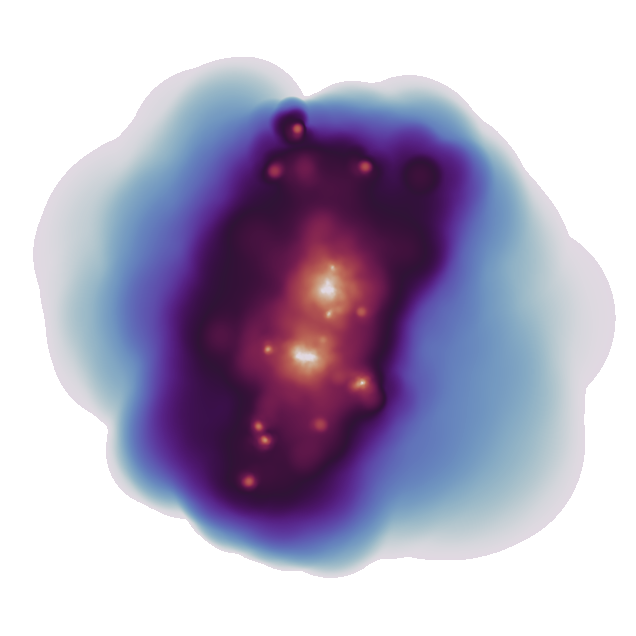}
\put (0,90) {\textcolor{black}{\textbf{L=27}}}
\end{overpic}
    \includegraphics[width=.81in]{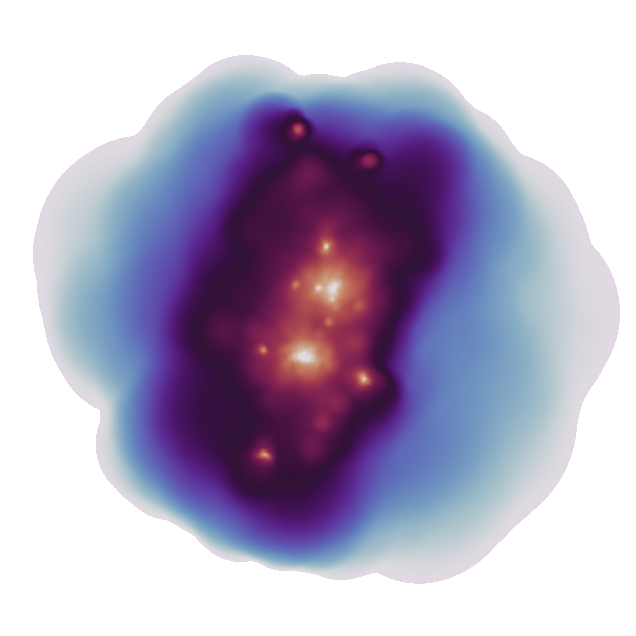}
    \includegraphics[width=.81in]{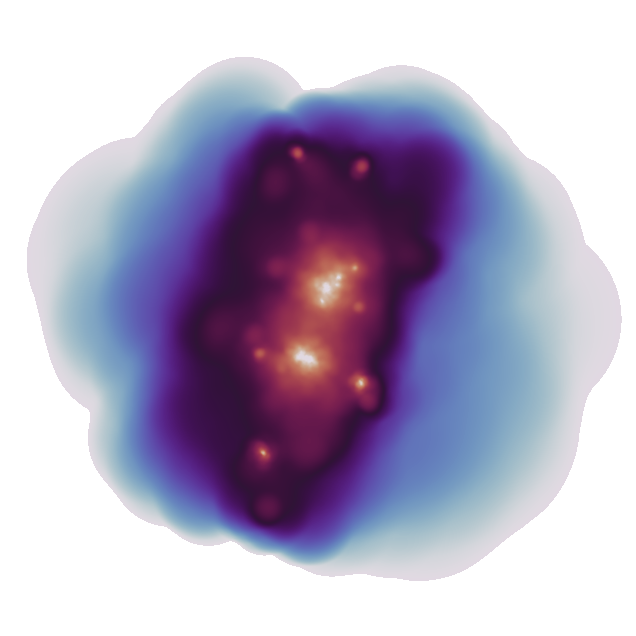}
    \includegraphics[width=.81in]{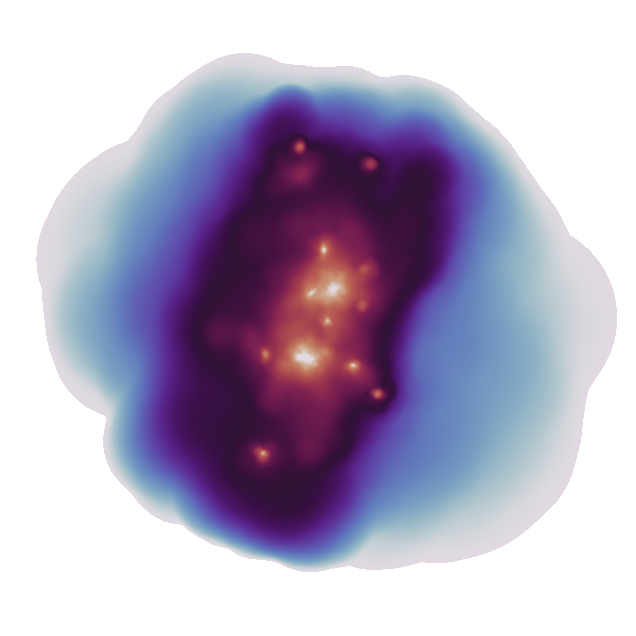}
\begin{overpic}[width=.81in]{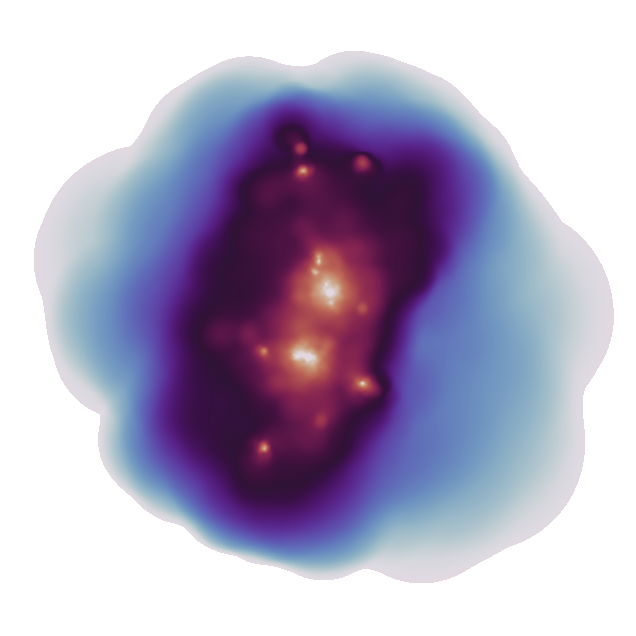}
\put (0,90) {\textcolor{black}{\textbf{L=28}}}
\put (0,0) {\textcolor{black}{\textbf{n=1}}}
\end{overpic}
\begin{overpic}[width=.81in]{plots/sphview_higher_L28_1.png}
\put (0,0) {\textcolor{black}{\textbf{n=2}}}
\end{overpic}
\begin{overpic}[width=.81in]{plots/sphview_higher_L28_1.png}
\put (0,0) {\textcolor{black}{\textbf{n=3}}}
\end{overpic}
\begin{overpic}[width=.81in]{plots/sphview_higher_L28_1.png}
\put (0,0) {\textcolor{black}{\textbf{n=4}}}
\end{overpic}
    \caption{The effect of varying the phase information on levels 25 ($\lc = 0.4$~cMpc, top row), 26 ($\lc = 0.2$~cMpc, second row), 27 ($\lc = 0.1$~cMpc, third row) and 28 ($\lc = 0.05$~cMpc, bottom row) on the projected density of LG candidate $[i=24943, j=366]$. }
    \label{fig:shift-comparison}
\end{figure}

\begin{figure}
    \includegraphics[width=\columnwidth]{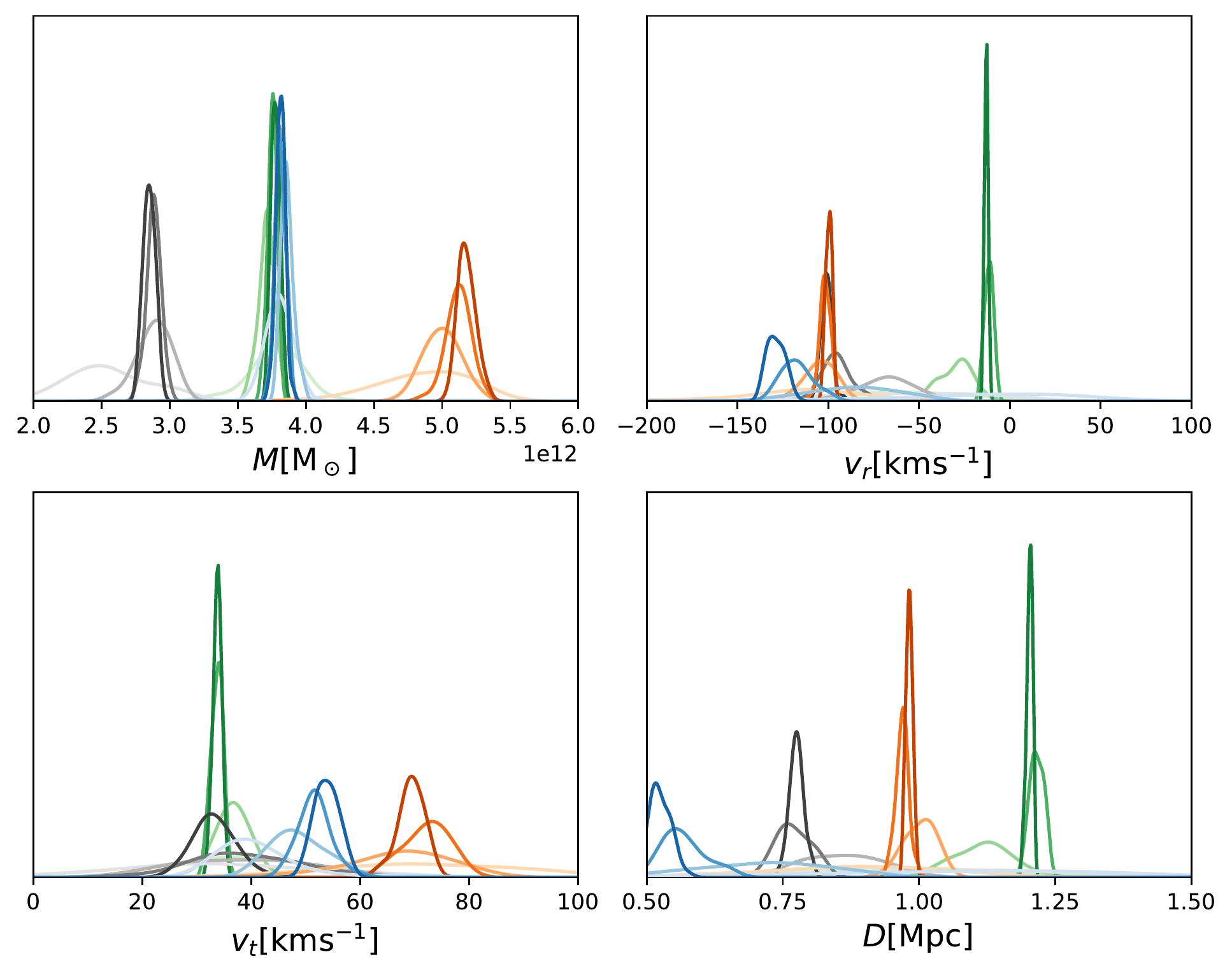}
    \caption{Probability density functions of LG observables; mass, radial velocity, transverse velocity and separation for smaller scale variations of four different Local Group variants, [i=21826, j=1] (greens), [i=41800, j=1] (oranges), [i=43595, j=1] (blues) and [i=24943, j=366] (greys), analogous to Figure~\ref{fig:stats:1D}. From light to dark, four different shades show variations at levels 25, 26, 27 and 28 ($\lc = 0.4, 0.2, 0.1, 0.05$~cMpc, respectively). Reducing the scale of the variations in the initial conditions reduces the scatter in all observed variables. Note that, as the PDFs are independently normalised for each variable, the height of the curves can be compared for different variations on the same panel, but not across panels.}
    \label{fig:stats:1D-higher}
\end{figure}

\section{Smaller Scale Variations} \label{sec:higher-order}
By varying higher levels of the octree, we can create even smaller scale variations of existing Local Group analogues. From the 0.8~cMpc sets, we selected four candidates, $[i=21868, j = 1]$, $[i=41800, j=1]$, $[i=43595, j=1]$ and $[i=24943, j=366]$, the latter of which yielded a LG that matches the observations. For each one, we created four new sets of 50 variations each, by varying the phase information at and above levels 25, 26, 27 and 28, corresponding to $\lc=0.4$, 0.2, 0.1, and 0.05~cMpc, respectively.

In Figure~\ref{fig:shift-comparison}, we illustrate the effect for [i=24943, j=366]. The four columns show the effects of applying four different shifts in {\sc Panphasia} at levels 25, 26, 27 and 28 of the octree (top to bottom). All variations shown contain a LG analogue, and as expected, the variation among them decreases from top to bottom with decreasing scale of the variation in the density field. At $\LL \geq 26$, even some small individual substructures can be easily identified across the different variants.

\subsection{Constraining the MW-M31 orbital phase} \label{sec:orbital-phase}
Figure~\ref{fig:stats:1D-higher} shows the probability density functions of the mass, radial velocity, transverse velocity and separation for 50 variations each at levels 25, 26, 27 and 28 ($\lc = 0.4, 0.2, 0.1, 0.05$, respectively) of the four Local Group variants from the 0.8~cMpc set listed above. The scatter in all observables decreases as the scale of the variations is decreased.

It is worth noting that at $\LL=25$, not all simulations have yielded a LG analogue that matches the "loose" criteria. Variations at $\LL=25$ of [i=21826,j=1] contain a LG analogue in 47/50 cases. For [i=41800, j=1], [i=24943, j=366] and [i=43595, j=1], the rate is 49/50, 44/50 and 41/50, respectively. Higher level variations of [i=21826,j=1], [i=41800,j=1] and [i=24943, j=366] always contain a LG that fits at least the loose criteria, but for [i=43595, j=1], the number remains around 40--45: the separation is very close to 500 kpc, and a fraction of LG analogues fall below the threshold. %CHECK

\begin{figure}
    \includegraphics[width=\columnwidth]{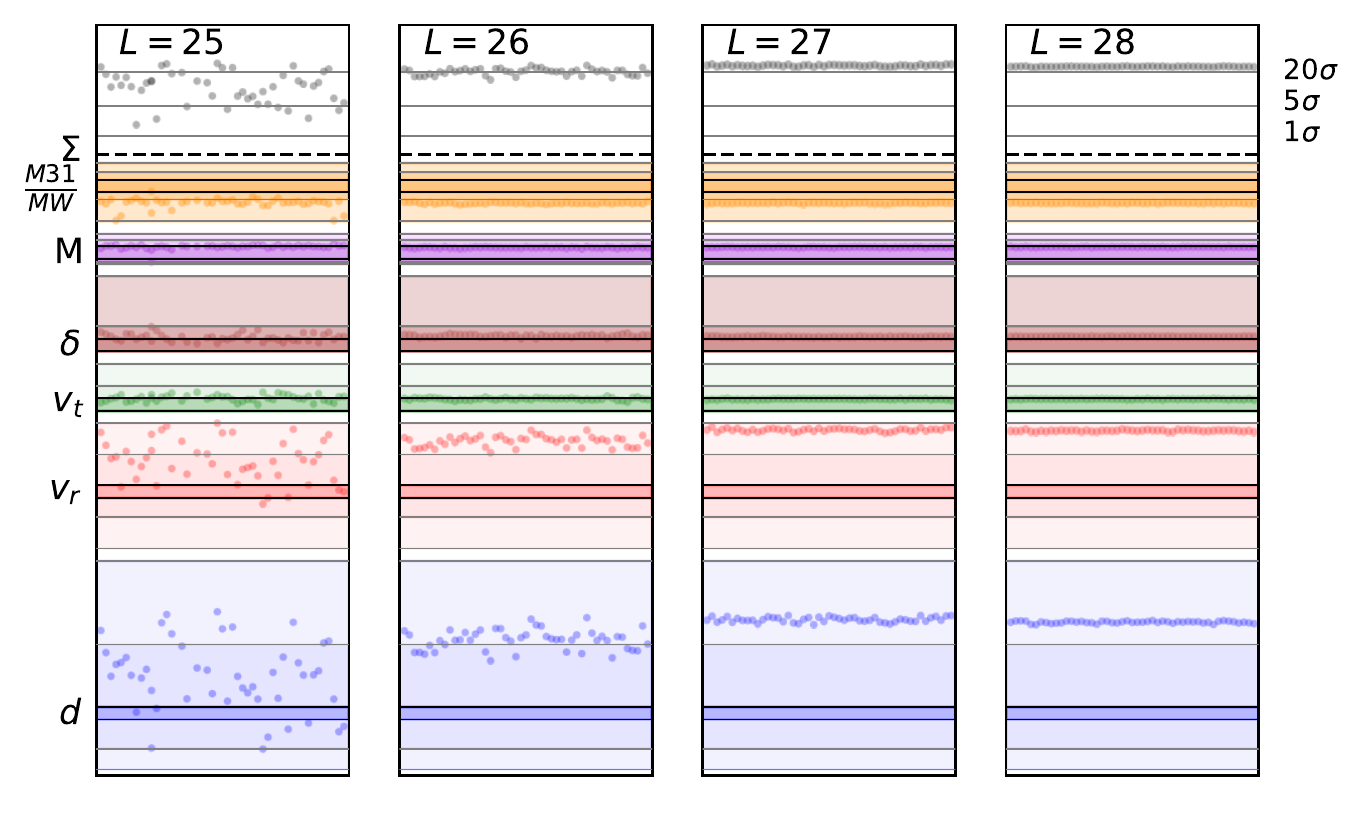}
    
    \vspace{-2mm} 
    
    \includegraphics[width=\columnwidth]{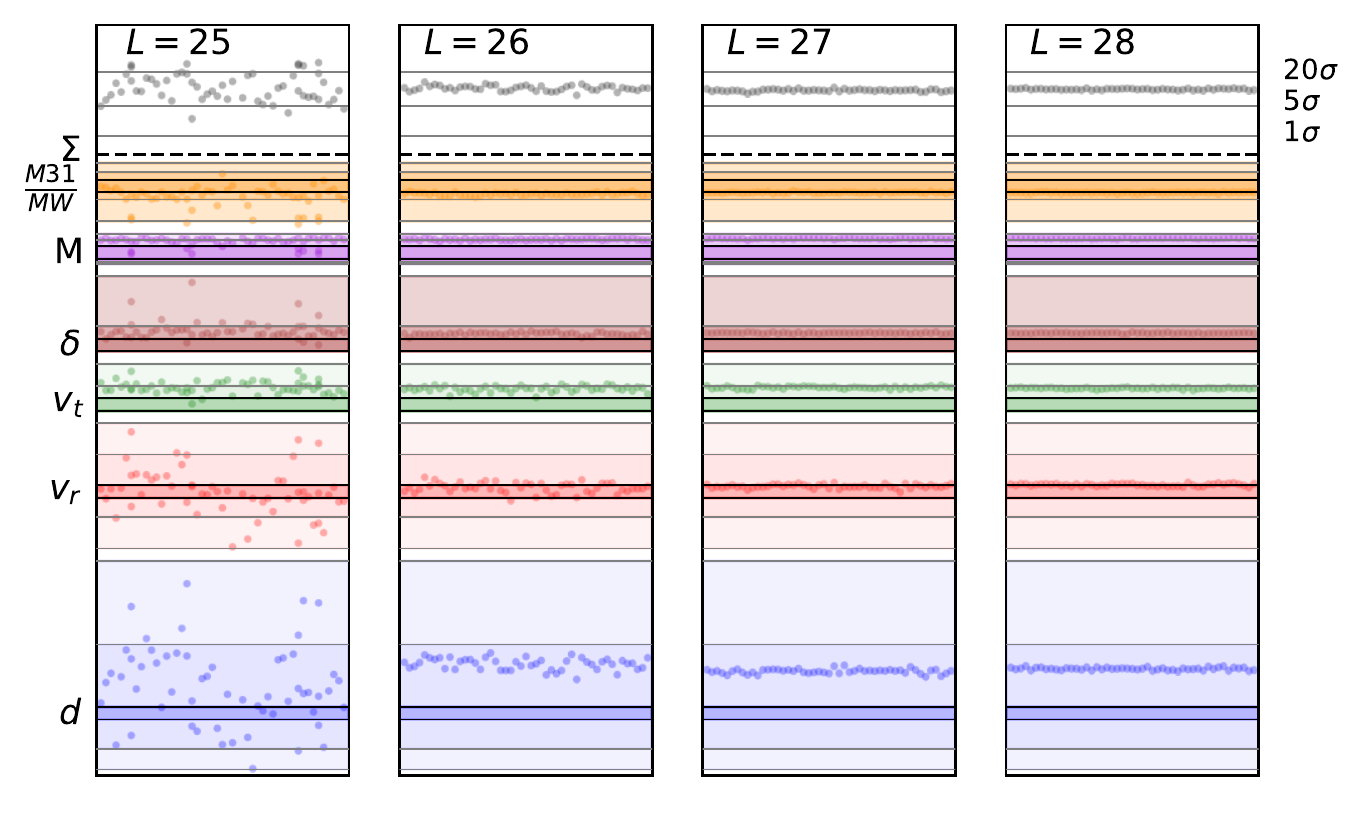}
     
    \vspace{-2mm}
    
    \includegraphics[width=\columnwidth]{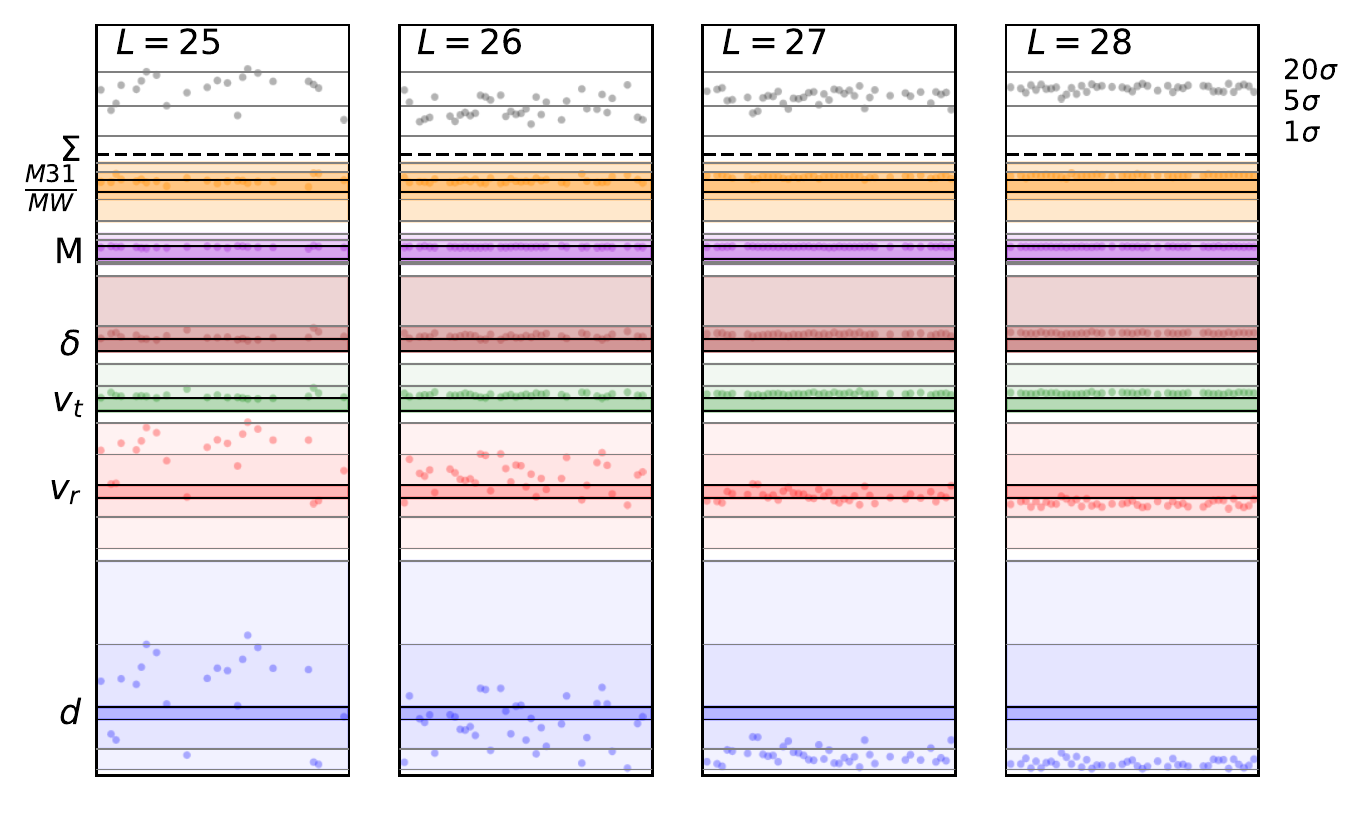}
     
    \vspace{-2mm} 
     
    \includegraphics[width=\columnwidth]{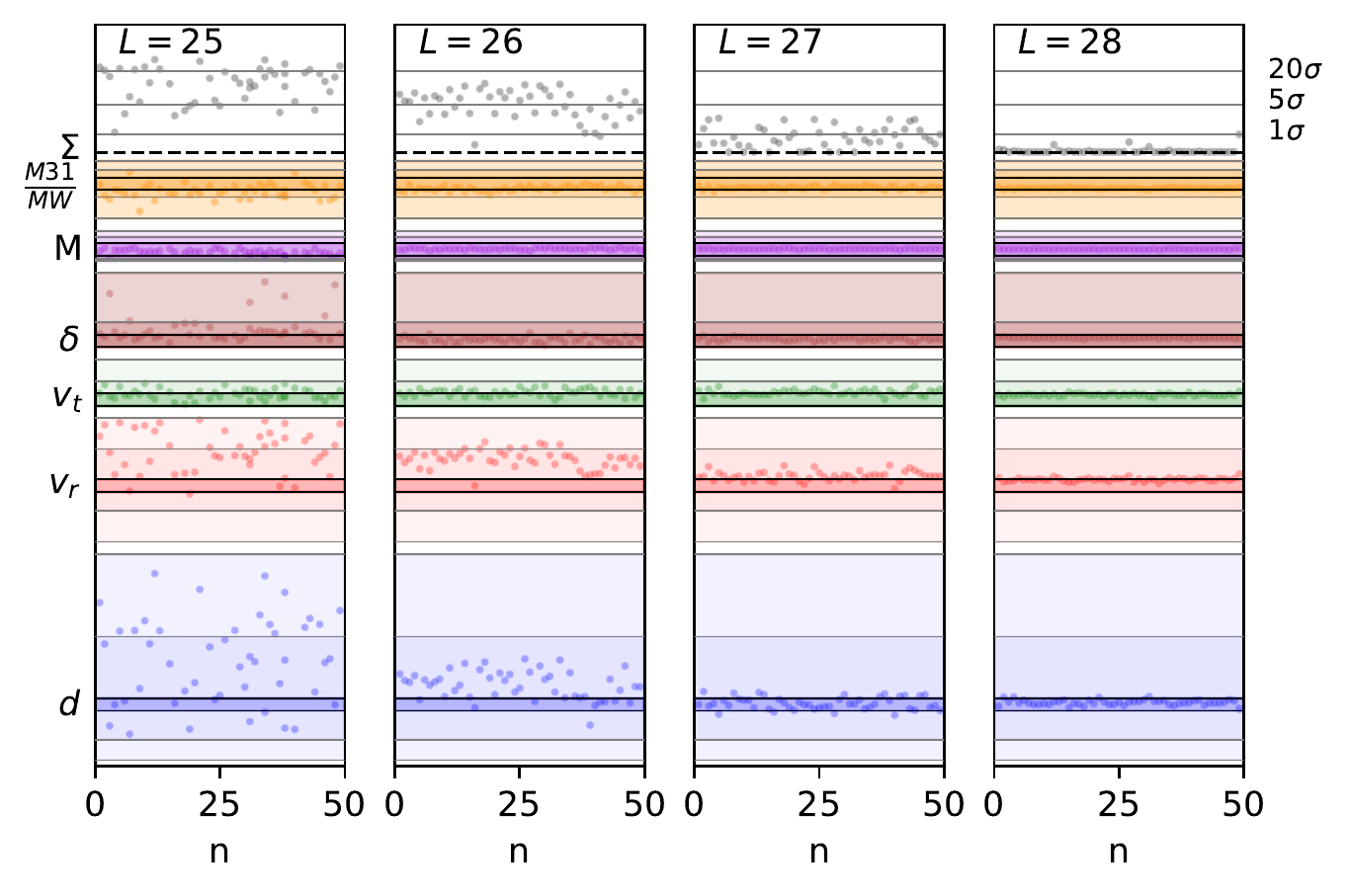}
    
    \vspace{-1mm} 
    
    \caption{Quality plots, analogous to Figures~\ref{fig:quality-first-order} and ~\ref{fig:quality-second-order}, of higher order variations at levels 25, 26, 27 and 28 ($\lc = 0.4, 0.2, 0.1, 0.05$~cMpc, respectively) of four existing variants, [i=21826, j=1] (top row), [i=41800, j=1] (second row), [i=43595, j=1] (third row) and [i=24943, j=366] (bottom row),  In all cases, the scatter in observables decreases further with decreasing scale of variation. By level 27, the variances in separation and radial velocity become comparable to the observational uncertainties. At this scale of variation, the phase of the orbit is set. 
    }
    \label{fig:quality-higher-order}
\end{figure}

The extent to which individual variables vary at each level depends on their values. For example, the scatter in separation, $d$, is largest for the sets that have the {\it lowest} separation. Conversely, the scatter in radial velocity is largest for the sets that have the most negative velocity and smallest for velocities close to zero. This can be readily understood if, as explained in Section~\ref{sec:characteristics}, the small-scale variations are, to a large extent, perturbations to the phase of the Local Group's orbit, whose integrals of motion have already been defined on larger scales. Local Group variants with large separation are close to apocentre, where the orbital velocity is low. In this case, a perturbation to the phase causes only a small change in position and velocity. On the other hand, Local Group variants with small separation are closer to pericentre, and the same perturbation to the phase causes a larger change in position and velocity.

To determine which scales in the primordial density field of the Local Universe were responsible for setting the phase of the Local Group's orbit, we are thus primarily interested in perturbations close to the observed values, i.e. $d\sim 770$~kpc, $v_r \sim -110$~kms$^{-1}$. In Table~\ref{tab:sensitivity}, we list the sensitivity of different parameters to the variations of LG candidate [i=24943, j=366].

\begin{table}
    \makegapedcells
    \centering
    \begin{tabular}{l|c|c|c|c|c}
        L &  $\lc$ & $M \pm \sigma_{M}$ & $d \pm \sigma_d$ & $v_r \pm \sigma_{v_r}$ & $v_t \pm \sigma_{v_t}$ \\
%           \midrule 
        & [cMpc] &  $[10^{12}\Ms] $ & [kpc] & [kms$^{-1}$] & [kms$^{-1}$] \\
        \midrule
        25 & 0.4  & $2.53 \pm 0.28$ & $1044 \pm 205$ & $-57 \pm 36$ & $30 \pm 16$ \\
        26 & 0.2  & $2.88 \pm 0.12$ & $857 \pm 68$   & $-67 \pm 14$ & $9 \pm 11$  \\ 
        27 & 0.1  & $2.88 \pm 0.05$ & $766 \pm 29$   & $-96 \pm 8$ & $37 \pm 9$  \\
        28 & 0.05 & $2.85 \pm 0.04$ & $774 \pm 12$   & $-101 \pm 3$ & $33 \pm 4$ \\
    \end{tabular}
    \caption{Sensitivity of the Local Group's observables to small-scale variations of the initial density for candidate [i=24943, j=366]. L is the level of the octree on which coefficients are randomised, $\lc$ is the corresponding cut-off scale for the $\Lambda$CDM power spectrum. M, $d$, $v_r$ and $v_t$ are the median values of the LG mass, separation, radial velocity and tangential velocity, respectively, while $\sigma_M$, $\sigma_d$, $\sigma_{v_r}$ and $\sigma_{v_t}$ are the 1-sigma scatter in the corresponding observables.}
    \label{tab:sensitivity}
\end{table}

For all scales shown, the mass of the LG varies by $\sim 10\%$ or less, well below the observational uncertainty. At $\LL=25$, the scatter in separation and radial velocity is $\sim 200$~kpc and $\sim 40$kms$^{-1}$, respectively. It is worth noting that the median separation and median radial velocity are those of a LG close to apocentre, resulting in smaller variance.

The variance in separation and radial velocity become comparable to the observational uncertainty only for levels $\LL=27$. While the precise threshold appears somewhat arbitrary, we conclude that all scales down to $\LL=26$, or  $\lc=0.2$~cMpc, have had a significant impact on setting the Local Group's orbital phase, and its current relative velocity and separation. For comparison, in \cite{Sawala-2020}, we had computed the average scatter in the position and velocity of central haloes of mass $\sim 10^{12}\;\Ms$ to be $\sim 20 $~kpc, and $\sim 3$~km~s$^{-1}$, respectively, from perturbations of the same scale. That these values are somewhat lower than those reported here shows that the relative positions and velocity of the MW and M31 are more sensitive to small-scale changes. The scale of $\lc=0.2$~cMpc is also sufficient to determine the existence of individual haloes down to $2\times 10^9\;\Ms$. In the hierarchy of scales that determine the formation of the LG, the scales that set the phase of the Local Group's orbit are thus significantly smaller than those which determine the presence of LMC/M33 analogues.

As is evident from Figures~\ref{fig:stats:1D-higher} and~\ref{fig:quality-higher-order}, the parameters of LG analogues initially chosen from the $0.8$~cMpc set are not necessarily at the centres of the different variations. Every LG variant contains phase information at all levels that we can resolve in our simulations, and represents just one random sample of the possible variations. When we consider perturbations to the orbital phase of a LG analogue, for individual observables, especially for Local Groups that are close to pericentre, the expected "reversion to the mean" becomes a "reversion to apocentre", where a larger fraction of the orbital phase is spent.

\section{Conclusion}\label{sec:conclusion}
Beginning with the {\sc BORG} reconstruction of the local large scale structure, we have constructed hybrid constrained initial conditions that embed the Local Group in its cosmic environment. Using the techniques described in \cite{Sawala-2020}, our simulations supplement the large scale constraints with independent randomised phase information below $\lc = 6.48$ cMpc within a 16~cMpc sphere around the observer. Depending on the selection criteria, we find many approximate LG analogues, and for most of them, we find several, and often many variants of the same object. This leads us to conclude that we have, most likely, explored a large fraction of the event space of LGs that can form with these particular constraints assuming $\Lambda$CDM.

We have identified the scales of the initial density field that determine the different Local Group properties. We find that the classical observables, such as the separation and radial velocities are not only transient in time, but also subject to very small scale changes in the initial density field: we find that the phase information needs to be fixed down to $\lc=0.2$~cMpc to constrain the phase of the orbit. On the other hand, as discussed in Section~\ref{sec:constants}, the orientation of the LG with respect to the large scale structure, the total mass, and integrals of motion such as the orbital energy and angular momentum, are set on larger scales ($\lc = 0.8$~cMpc). We suggest to consider these as the defining characteristics of the LG.

In constructing and refining cosmological initial conditions, this hierarchy of observables also has practical consequences. We find that in our simulations, the scatter in LG observables introduced by small-scale variations of the initial density field depends strongly on the presence of additional haloes. Furthermore, given our constraints on the larger scales, the scatter introduced by small scale variations is required in order to produce a Local Group with the observed properties. This in itself does not necessarily imply that the presence of the LMC / M33 is necessary for a MW-M31 pair to match the other observables. However, given how strongly the presence of such objects affects the orbit, and given that their formation is set on scales greater than those which set the orbital phase, it is clear that the actual Local Group, in its true cosmic environment, would not have the observed kinematics without the presence of the  LMC and M33. Consequently, simulations that produce a LG with a precise match to the other observables in the absence of those additional massive objects, would miss important aspects of its environment.

The current work uses dark matter only simulations. Given the level of detail we require to identify LG objects, this is justified, but using hydrodynamic simulations would allow us to select also by galaxy properties. In principle, there is nothing other than computational cost that would prevent us from employing hydrodynamic simulations in our hierarchical method for constructing initial conditions.

In this work, we have been interested in exploring the sensitivity of observables to small scale variations of the initial density field. Of course, in constructing higher level variations, it would be possible to select specific candidates at each level for refinement at smaller scales, and to create ever more precise matches to the observed Local Group. We will fully explore smaller scale variations, and their impact on substructures, in a forthcoming paper. %There is indeed plenty of room at the bottom.

All constrained simulations of the Local Universe should eventually converge to the same result. One major caveat in our current work is the residual variance in the different realisations of the local large scale structure.However, this may only be a temporary limitation. In principle, our method of iteratively randomising smaller scales allows us to explore all possible Local Groups that can form in the Local Universe, and to do so in a very efficient way. If the $\Lambda$CDM model was manifestly incorrect, we might not have found a precise LG analogue within the constraints used. With improving constraints, we now have a clear pathway towards not only creating Local Group analogues, but eventually reproducing {\it the} Local Group. Alternatively, our method would also allow us to demonstrate that the Local Group cannot possibly form given a set of large scale constraints and cosmological parameters. We envisage this to become a powerful and comprehensive way of testing cosmological and astrophysical assumptions.

\section*{Data Availability Statement}
The data underlying this article will be shared on reasonable request to the corresponding author.

\section*{Acknowledgements}
TS is an Academy of Finland Research fellow. TS and SM are supported by the Academy of Finland grant 314238. PHJ acknowledges the support by the European Research Council via ERC Consolidator Grant (no. 818930).  CSF acknowledges support from European Research Council (ERC) Advanced Investigator grant DMIDAS (GA 786910). This work was also supported by the Consolidated Grant for Astronomy at Durham (ST/L00075X/1). GL acknowledges financial support from the ANR BIG4, under reference ANR-16-CE23-0002. JJ acknowledges support by the Swedish Research Council (VR) under the project 2020-05143 -- "Deciphering the Dynamics of Cosmic Structure".

This work used the DiRAC@Durham facility managed by the Institute for Computational Cosmology on behalf of the STFC DiRAC HPC Facility (www.dirac.ac.uk). The equipment was funded by BEIS capital funding via STFC capital grants ST/K00042X/1, ST/P002293/1 and ST/R002371/1,
Durham University and STFC operations grant ST/R000832/1. DiRAC is part of the National e-Infrastructure. This work also used facilities hosted by the CSC - IT Centre for Science, Finland. We gratefully acknowledge the use of open source software, including Py-SPHViewer \citep{pysph-paper}, Matplotlib \citep{matplotlib-paper} and NumPy \citep{numpy-paper}.

%%%%%%%%%%%%%%%%%%%%%%%%%%%%%%%%%%%%%%%%%%%%%%%%%%

%%%%%%%%%%%%%%%%%%%% REFERENCES %%%%%%%%%%%%%%%%%%

\bibliographystyle{mnras} \bibliography{paper}

%%%%%%%%%%%%%%%%%%%%%%%%%%%%%%%%%%%%%%%%%%%%%%%%%%

%%%%%%%%%%%%%%%%% APPENDICES %%%%%%%%%%%%%%%%%%%%%
\bsp	% typesetting comment
\label{lastpage}
\end{document}